%% file: main.tex
\pdfoutput=1

\documentclass[12pt,a4paper]{article}

\usepackage{ifthen} 
\newboolean{pdflatex}
\setboolean{pdflatex}{true} 

\newboolean{articletitles}
\setboolean{articletitles}{true} 

\newboolean{uprightparticles}
\setboolean{uprightparticles}{false} 

\def\paperauthors{LHCb collaboration} 
\def\paperasciititle{Observation of several sources of CP violation in B+ -> K+ pi+ pi- decays} 
\def\papertitle{Observation of several sources of $C\!P$ violation in $B^+ \!\to K^+ \pi^+ \pi^-$ decays} 
\def\paperkeywords{{High Energy Physics}, {LHCb}} 
\def\papercopyright{\the\year\ CERN for the benefit of the LHCb collaboration} 
\def\paperlicence{CC BY 4.0 licence}
\def\paperlicenceurl{https://creativecommons.org/licenses/by/4.0/}

\newif\ifEnableSectionTOCLinks
\EnableSectionTOCLinksfalse 

\input{LHCb/preamble}
\usepackage{longtable} 
\usepackage{float}

\begin{document}

\renewcommand{\thefootnote}{\fnsymbol{footnote}}
\setcounter{footnote}{1}

\input{title-LHCb-PAPER}

\renewcommand{\thefootnote}{\arabic{footnote}}
\setcounter{footnote}{0}


\cleardoublepage


\pagestyle{plain} 
\setcounter{page}{1}
\pagenumbering{arabic}


\input{body}

\input{LHCb/acknowledgements}

\newpage
\input{appendix}
\clearpage

\addcontentsline{toc}{section}{References}
\bibliographystyle{LHCb/LHCb}
\bibliography{main,LHCb/standard,LHCb/LHCb-PAPER,LHCb/LHCb-CONF,LHCb/LHCb-DP,LHCb/LHCb-TDR}

\newpage
\input{Authorship_LHCb-PAPER-2025-069}

\end{document}

%% file: LHCb/preamble.tex
\usepackage[top=1in, bottom=1.25in, left=1in, right=1in]{geometry}

\usepackage{microtype}
\usepackage{lineno}  
\usepackage{xspace} 
\usepackage{caption} 

\usepackage{graphicx}  
\usepackage{color}
\usepackage{colortbl}
\graphicspath{{./figs/}} 

\usepackage{amsmath} 
\usepackage{amssymb}
\usepackage{amsfonts}
\usepackage{upgreek} 

\usepackage[normalem]{ulem} 

\newcommand*\patchAmsMathEnvironmentForLineno[1]{%
\expandafter\let\csname old#1\expandafter\endcsname\csname #1\endcsname
\expandafter\let\csname oldend#1\expandafter\endcsname\csname
end#1\endcsname
 \renewenvironment{#1}%
   {\linenomath\csname old#1\endcsname}%
   {\csname oldend#1\endcsname\endlinenomath}%
}
\newcommand*\patchBothAmsMathEnvironmentsForLineno[1]{%
  \patchAmsMathEnvironmentForLineno{#1}%
  \patchAmsMathEnvironmentForLineno{#1*}%
}
\AtBeginDocument{%
\patchBothAmsMathEnvironmentsForLineno{equation}%
\patchBothAmsMathEnvironmentsForLineno{align}%
\patchBothAmsMathEnvironmentsForLineno{flalign}%
\patchBothAmsMathEnvironmentsForLineno{alignat}%
\patchBothAmsMathEnvironmentsForLineno{gather}%
\patchBothAmsMathEnvironmentsForLineno{multline}%
\patchBothAmsMathEnvironmentsForLineno{eqnarray}%
}

\usepackage[pdftex,
            pdfauthor={\paperauthors},
            pdftitle={\paperasciititle},
            pdfkeywords={\paperkeywords}]{hyperref}
\usepackage{hyperxmp}
\hypersetup{
    pdfcopyright={Copyright (C) \papercopyright},
    pdflicenseurl={\paperlicenceurl}
}

\usepackage[colorinlistoftodos,textsize=scriptsize]{todonotes}

\usepackage[bottom,flushmargin,hang,multiple]{footmisc}

\usepackage[all]{hypcap} 

\input{LHCb/lhcb-symbols-def} 

\hypersetup{
  colorlinks   = true, 
  urlcolor     = blue, 
  linkcolor    = blue, 
  citecolor    = red   
}

\ifEnableSectionTOCLinks
    \usepackage[explicit]{titlesec} 
    
    \let\oldcontentsline\contentsline
    \renewcommand

    \titleformat{\section}{\normalfont\Large\bf}{\hyperlink{tocsection.\thesection}{{\thesection} \parbox[t]{\dimexpr\textwidth-1pc}{#1}}}{1pc}{}

    \titleformat{\subsection}{\normalfont\bf}{\hyperlink{tocsubsection.\thesubsection}{{\thesubsection} \parbox[t]{\dimexpr\textwidth-1pc}{#1}}}{1pc}{}

    \titleformat{name=\section,numberless}[display]{}{}{0pt}{\normalfont\Huge\bfseries #1}
\fi

\usepackage{cite} 
\usepackage{LHCb/mciteplus}
\makeatletter
\g@addto@macro\bfseries{\boldmath}
\makeatother

%% file: LHCb/lhcb-symbols-def.tex
\usepackage{xspace} 
\usepackage{upgreek}

\def\lhcb   {\mbox{LHCb}\xspace}

\def\MagUp {\mbox{\em Mag\kern -0.05em Up}\xspace}

\ifdefined\ifuprightparticles
\else
\newboolean{uprightparticles}
\setboolean{uprightparticles}{false} 
\fi

\ifthenelse{\boolean{uprightparticles}}%
{

 \def\Ppi         {\ensuremath{\uppi}\xspace}

 \def\Ppsi        {\ensuremath{\uppsi}\xspace}

 \def\PDelta      {\ensuremath{\Delta}\xspace}                 
 \def\PXi         {\ensuremath{\Xi}\xspace}                 
 \def\PLambda     {\ensuremath{\Lambda}\xspace}                 
 \def\PSigma      {\ensuremath{\Sigma}\xspace}                 
 \def\POmega      {\ensuremath{\Omega}\xspace}                 
 \def\PUpsilon    {\ensuremath{\Upsilon}\xspace}
 \let\oldPi\Pi
 \def\PPi         {\ensuremath{\oldPi}\xspace}

 \def\PB      {\ensuremath{\mathrm{B}}\xspace}                 
 \def\PD      {\ensuremath{\mathrm{D}}\xspace}                 
 \def\PJ      {\ensuremath{\mathrm{J}}\xspace}                 
 \def\PK      {\ensuremath{\mathrm{K}}\xspace}                 
 \def\Pb      {\ensuremath{\mathrm{b}}\xspace}                 
 \def\Pc      {\ensuremath{\mathrm{c}}\xspace}                 
 \def\Pd      {\ensuremath{\mathrm{d}}\xspace}                 

 \def\Pp      {\ensuremath{\mathrm{p}}\xspace}                 

 \def\Ps      {\ensuremath{\mathrm{s}}\xspace}                 
                  
 \def\Pu      {\ensuremath{\mathrm{u}}\xspace}                 
 \def\thebaroffset{0.0em}
}
{

 \def\Ppi         {\ensuremath{\pi}\xspace}

 \def\Ppsi        {\ensuremath{\psi}\xspace}                 
                  
 \mathchardef\PDelta="7101
 \mathchardef\PXi="7104
 \mathchardef\PLambda="7103
 \mathchardef\PSigma="7106
 \mathchardef\POmega="710A
 \mathchardef\PUpsilon="7107
 \mathchardef\PPi="7105
 \def\PB      {\ensuremath{B}\xspace}                 
 \def\PD      {\ensuremath{D}\xspace}                 
 \def\PJ      {\ensuremath{J}\xspace}                 
 \def\PK      {\ensuremath{K}\xspace}                 
 \def\Pb      {\ensuremath{b}\xspace}                 
 \def\Pc      {\ensuremath{c}\xspace}                 
 \def\Pd      {\ensuremath{d}\xspace}                 

 \def\Pp      {\ensuremath{p}\xspace}                 

 \def\Ps      {\ensuremath{s}\xspace}                 
                  
 \def\Pu      {\ensuremath{u}\xspace}                 
 \def\thebaroffset{0.18em}
}
\newcommand{\offsetoverline}[2][\thebaroffset]{\kern #1\overline{\kern -#1 #2}}%

\makeatletter
\ifcase \@ptsize \relax
  \newcommand{\miniscule}{\@setfontsize\miniscule{4}{5}}
\or
  \newcommand{\miniscule}{\@setfontsize\miniscule{5}{6}}
\or
  \newcommand{\miniscule}{\@setfontsize\miniscule{5}{6}}
\fi
\makeatother

\DeclareRobustCommand{\optbar}[1]{\shortstack{{\miniscule (\rule[.5ex]{1.25em}{.18mm})}
  \\ [-.7ex] $#1$}}

\def\uquark    {{\ensuremath{\Pu}}\xspace}

\def\dquark    {{\ensuremath{\Pd}}\xspace}

\def\squark    {{\ensuremath{\Ps}}\xspace}

\def\cquark    {{\ensuremath{\Pc}}\xspace}
\def\cquarkbar {{\ensuremath{\overline \cquark}}\xspace}

\def\bquark    {{\ensuremath{\Pb}}\xspace}

\def\pion   {{\ensuremath{\Ppi}}\xspace}

\def\pip    {{\ensuremath{\pion^+}}\xspace}
\def\pim    {{\ensuremath{\pion^-}}\xspace}

\def\kaon    {{\ensuremath{\PK}}\xspace}
\def\Kbar    {{\ensuremath{\offsetoverline{\PK}}}\xspace}

\def\KorKbar {\kern \thebaroffset\optbar{\kern -\thebaroffset \PK}{}\xspace}

\def\Kp      {{\ensuremath{\kaon^+}}\xspace}
\def\Km      {{\ensuremath{\kaon^-}}\xspace}

\def\Dbar    {{\ensuremath{\offsetoverline{\PD}}}\xspace}
\def\D       {{\ensuremath{\PD}}\xspace}

\def\DorDbar {\kern \thebaroffset\optbar{\kern -\thebaroffset \PD}\xspace}

\def\Dzb     {{\ensuremath{\Dbar{}^0}}\xspace}
\def\Dp      {{\ensuremath{\D^+}}\xspace}
\def\Dm      {{\ensuremath{\D^-}}\xspace}

\def\DpDm    {\ensuremath{\Dp {\kern -0.16em \Dm}}\xspace}

\def\B       {{\ensuremath{\PB}}\xspace}

\def\BorBbar {\kern \thebaroffset\optbar{\kern -\thebaroffset \PB}\xspace}
\def\Bz      {{\ensuremath{\B^0}}\xspace}

\def\Bd      {{\ensuremath{\B^0}}\xspace}

\def\BdorBdbar {\kern \thebaroffset\optbar{\kern -\thebaroffset \Bd}\xspace}
\def\Bu      {{\ensuremath{\B^+}}\xspace}
\def\Bub     {{\ensuremath{\B^-}}\xspace}
\def\Bp      {{\ensuremath{\Bu}}\xspace}
\def\Bm      {{\ensuremath{\Bub}}\xspace}
\def\Bpm     {{\ensuremath{\B^\pm}}\xspace}

\def\Bs      {{\ensuremath{\B^0_\squark}}\xspace}

\def\BsorBsbar {\kern \thebaroffset\optbar{\kern -\thebaroffset \Bs}\xspace}

\def\jpsi     {{\ensuremath{{\PJ\mskip -3mu/\mskip -2mu\Ppsi}}}\xspace}

\def\Y#1S{\ensuremath{\PUpsilon{(#1S)}}\xspace}

\def\proton      {{\ensuremath{\Pp}}\xspace}

\def\LorLbar     {\kern \thebaroffset\optbar{\kern -\thebaroffset \PLambda}\xspace}

\newcommand{\decay}[2]{\ensuremath{\mathinner{#1\!\to #2}}\xspace}

\def\to                 {\ensuremath{\rightarrow}\xspace}

\def\CP                {{\ensuremath{C\!P}}\xspace}

\def\Vud  {{\ensuremath{V_{\uquark\dquark}^{\phantom{\ast}}}}\xspace}
\def\Vcd  {{\ensuremath{V_{\cquark\dquark}^{\phantom{\ast}}}}\xspace}

\def\Vubs  {{\ensuremath{V_{\uquark\bquark}^\ast}}\xspace}
\def\Vcbs  {{\ensuremath{V_{\cquark\bquark}^\ast}}\xspace}

\def\AT#1     {\ensuremath{A_{\mathrm{T}}^{#1}}\xspace}           

\def\C#1      {\ensuremath{\mathcal{C}_{#1}}\xspace}                       
\def\Cp#1     {\ensuremath{\mathcal{C}_{#1}^{'}}\xspace}                    
\def\Ceff#1   {\ensuremath{\mathcal{C}_{#1}^{\mathrm{(eff)}}}\xspace}        
\def\Cpeff#1  {\ensuremath{\mathcal{C}_{#1}^{'\mathrm{(eff)}}}\xspace}       
\def\Ope#1    {\ensuremath{\mathcal{O}_{#1}}\xspace}                       
\def\Opep#1   {\ensuremath{\mathcal{O}_{#1}^{'}}\xspace}                    

\newcommand{\aunit}[1]{\ensuremath{\text{\,#1}}}       

\newcommand{\tev}{\aunit{Te\kern -0.1em V}\xspace}
\newcommand{\gev}{\aunit{Ge\kern -0.1em V}\xspace}
\newcommand{\mev}{\aunit{Me\kern -0.1em V}\xspace}
\newcommand{\kev}{\aunit{ke\kern -0.1em V}\xspace}
\newcommand{\ev}{\aunit{e\kern -0.1em V}\xspace}
 
\newcommand{\mevc}{\ensuremath{\aunit{Me\kern -0.1em V\!/}c}\xspace}
\newcommand{\gevc}{\ensuremath{\aunit{Ge\kern -0.1em V\!/}c}\xspace}
\newcommand{\mevcc}{\ensuremath{\aunit{Me\kern -0.1em V\!/}c^2}\xspace}
\newcommand{\gevcc}{\ensuremath{\aunit{Ge\kern -0.1em V\!/}c^2}\xspace}

\def\fb   {\ensuremath{\aunit{fb}}\xspace}
\def\invfb   {\ensuremath{\fb^{-1}}\xspace}

\def\gsim{{~\raise.15em\hbox{$>$}\kern-.85em
          \lower.35em\hbox{$\sim$}~}\xspace}
\def\lsim{{~\raise.15em\hbox{$<$}\kern-.85em
          \lower.35em\hbox{$\sim$}~}\xspace}

\newcommand{\Real}{\ensuremath{\mathcal{R}e}\xspace}

\def\tell1  {TELL1\xspace}
\def\ukl1   {UKL1\xspace}

\newcommand{\lhcborcid}[1]{\href{https://orcid.org/#1}{\hspace*{0.1em}\raisebox{-0.45ex}{\includegraphics[width=1em]{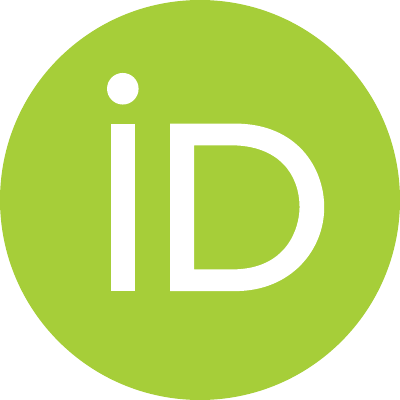}}}}


%% file: title-LHCb-PAPER.tex

\begin{titlepage}
\pagenumbering{roman}

\vspace*{-1.5cm}
\centerline{\large EUROPEAN ORGANIZATION FOR NUCLEAR RESEARCH (CERN)}
\vspace*{1.5cm}
\noindent
\begin{tabular*}{\linewidth}{lc@{\extracolsep{\fill}}r@{\extracolsep{0pt}}}
\ifthenelse{\boolean{pdflatex}}
{\vspace*{-1.5cm}\mbox{\!\!\!\includegraphics[width=.14\textwidth]{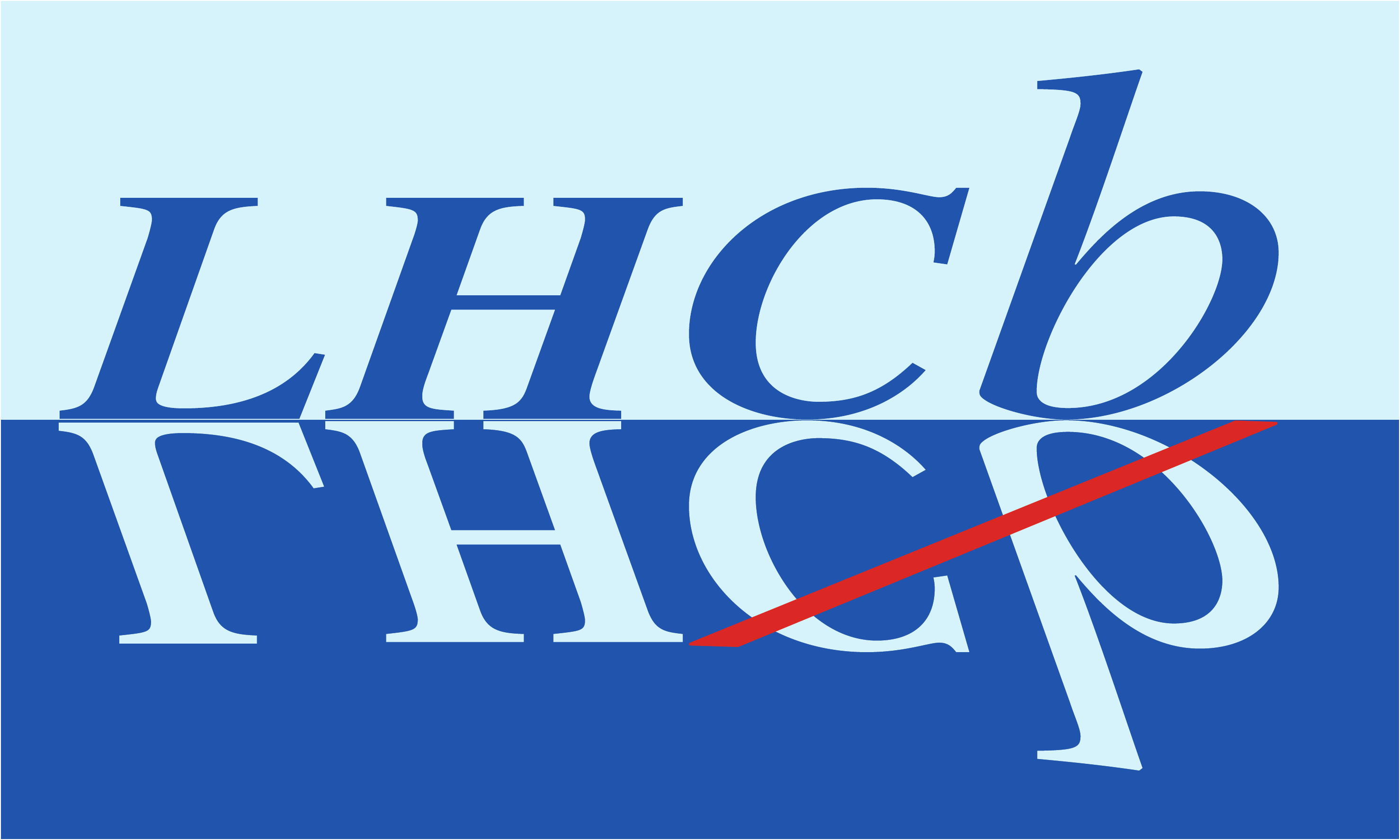}} & &}%
{\vspace*{-1.2cm}\mbox{\!\!\!\includegraphics[width=.12\textwidth]{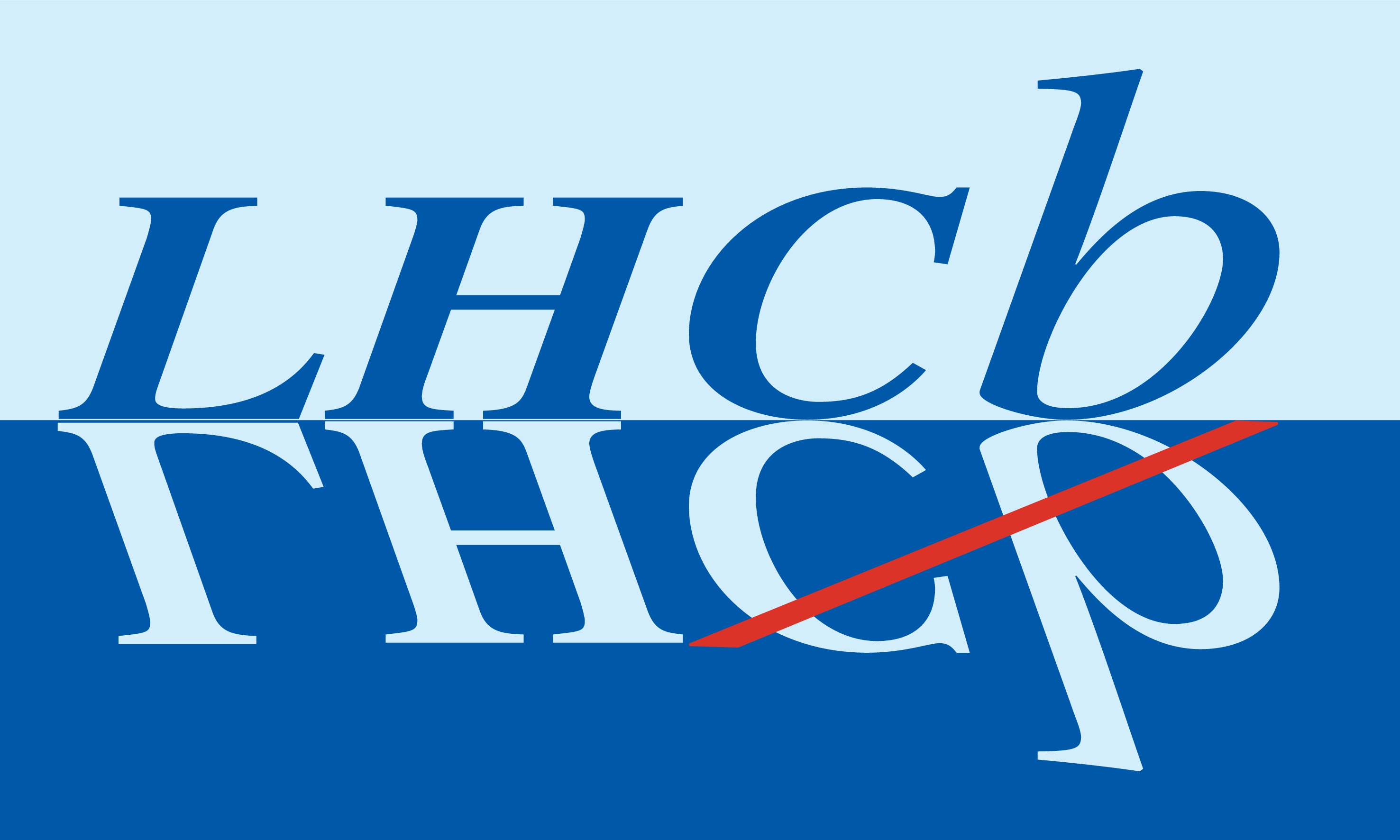}} & &}%
\\
 & & CERN-EP-2026-218 \\  
 & & LHCb-PAPER-2025-069 \\  
 & & 17 August 2026 \\ 
 & & \\
\end{tabular*}

\vspace*{4.0cm}

{\normalfont\bfseries\boldmath\huge
\begin{center}
  \papertitle 
\end{center}
}

\vspace*{2.0cm}

\begin{center}
\paperauthors\footnote{Authors are listed at the end of this Letter.}
\end{center}

\vspace{\fill}

\begin{abstract}
    \noindent
An amplitude analysis of $B^+ \!\to K^+ \pi^+ \pi^-$ decays is presented in which six $C\!P$-violating phenomena are judged to be of significance for the first time. This analysis is based on $pp$ collision data recorded with the LHCb detector in 2011--2012, corresponding to an integrated luminosity of $3\,\text{fb}^{-1}$. Quasi-two-body $C\!P$ violation in $B^+ \!\to \rho(770)^0 K^+$ decays is discovered, while $C\!P$ violation at amplitude level is established in $B^+ \!\to f_2(1270) K^+$ decays. First evidence for $C\!P$ violation is reported in both the fully elastic S-wave $\pi\pi$--$\pi\pi$ rescattering region and also for any decay involving a spin-3 resonance. Additionally, significant $C\!P$-violation effects are identified in the interference between different $\pi\pi$ partial waves, with observation in S-P wave interference and evidence in S-D wave interference, both of which must be driven by long-distance interactions.
\end{abstract}

\vspace*{2.0cm}

\begin{center}
  Submitted to
  Phys.~Rev.~Lett.
\end{center}

\vspace{\fill}

{\footnotesize 
\centerline{\copyright~\papercopyright. \href{\paperlicenceurl}{\paperlicence}.}}
\vspace*{2mm}

\end{titlepage}


\newpage
\setcounter{page}{2}
\mbox{~}

%% file: body.tex
Violation of the combined charge-conjugation and parity-transformation operations, \CP violation, gives rise to differences between the interactions of matter and antimatter particles.
In the Standard Model~(SM), \CP violation originates from a single irreducible complex phase in the Cabibbo–Kobayashi–Maskawa~(CKM) matrix that describes the strength of transitions between up- and down-type quarks in the weak interaction~\cite{Cabibbo:1963yz,Kobayashi:1973fv}.
While all measurements of \CP violation in particle decays to date appear consistent with this explanation, the degree of matter-antimatter asymmetry permitted by the SM through \CP violation is inconsistent with the macroscopic observed Universe~\cite{Shaposhnikov:1991cu}, motivating further studies and searches for sources of \CP violation beyond the SM.

For beauty hadrons, \CP violation in decay manifests as an asymmetry in particle and antiparticle decay rates, where large effects can arise when two or more different complex amplitudes contributing to a transition are of comparable strength. 
The phase of each amplitude can be decomposed into weak and strong components, which change sign and are invariant, respectively, under \CP\ transformations. 
Differences in both the weak and strong phases of the contributing amplitudes are required for an asymmetry in decay rates to occur. 
In $b$-hadron decay processes to charmless final states such as \decay{\Bp}{\Kp\pip\pim},\footnote{The inclusion of charge-conjugate processes is implied, except where asymmetries are discussed.} contributions from both $b \to u$ tree-level amplitudes and $b \to s$ or $b \to d$ loop (so-called ``penguin'') amplitudes are possible, and provide the required weak phase difference. 
Strong-phase differences may arise either from these transitions~\cite{Bander:1979px}, or through additional long-distance effects such as intermediate resonant structure and particle rescattering.

The \decay{\Bz}{\Kp\pim} decay provided the first observation of \CP violation in decay in the $b$-hadron system~\cite{BaBar:2004gyj,Belle:2004nch}, and large \CP-violation effects have been observed in regions of the Dalitz plots~\cite{Dalitz:1953cp,Fabri:1954zz} that describe the phase-space of three-body charmless \Bp\ decays~\cite{LHCb-PAPER-2013-027,LHCb-PAPER-2014-044,LHCb-PAPER-2021-049}.
Detailed amplitude models for \decay{\Bp}{\pip\pip\pim}~\cite{LHCb-PAPER-2019-018,LHCb-PAPER-2019-017} and \decay{\Bp}{\Kp\Km\pip}~\cite{LHCb-PAPER-2018-051} decays have been reported by the LHCb collaboration. For \decay{\Bp}{\Kp\pip\pim} and \decay{\Bp}{\Kp\Kp\Km} decays the only existing amplitude analyses are from the BaBar and Belle experiments~\cite{Belle:2005rpz,BaBar:2008lpx,BaBar:2012iuj,Belle:2004drb}, which did not have sufficient sensitivity to identify the origin of \CP-violation effects beyond those arising directly in quasi-two-body decays. 
The previous analyses of the \decay{\Bp}{\Kp\pip\pim} decay~\cite{Belle:2005rpz,BaBar:2008lpx} did, however, find evidence for \CP violation in the \decay{\Bp}{\rho(770)^0\Kp} process, and further support for \CP violation in the relevant region of the Dalitz plot has been found in a model-independent analysis that establishes a significant asymmetry in the $\pip\pim$ P-wave~\cite{LHCb-PAPER-2021-050}. 
Moreover, there are several theoretical  predictions for a sizeable \CP asymmetry in this process~\cite{Beneke:2003zv,Li:2006jv,Cheng:2009cn,Cheng:2014rfa,Bell:2015koa,Cheng:2016shb,Chai:2022ptk}, albeit with large  uncertainties. 
Theoretical predictions for \CP violation in \decay{\Bp}{f_2(1270)\Kp} decays~\cite{Cheng:2010yd,Li:2018lbd} are likewise in agreement with the previous experimental results.

Beyond the intrinsic interest in understanding the origin of \CP violation in these decays, improved results will provide pathways to test the SM. 
For example, within the SU(3) limit there are SM relations between the fully symmetrised \decay{\Bp}{\Kp\pip\pim} and \decay{\Bp}{\Kp\Kp\Km} decay amplitudes that can be tested experimentally~\cite{Bhattacharya:2014eca}. 
Including results from similar three-body charmless \Bd decays, these symmetrised amplitudes can be used to determine the fundamental \CP-violating phase of the CKM matrix, $\gamma = \phi_3 \equiv \arg[-\Vud \Vubs/\Vcd \Vcbs]$~\cite{Rey-LeLorier:2011ltd,Bhattacharya:2013cla,Bertholet:2018tmx}. 
A similar analysis can also be performed for the fully antisymmetric amplitudes~\cite{Bhattacharya:2023pef}.
These methods are sensitive to new heavy particles entering the $b \to s$ penguin loop, and the measured values of $\gamma$ could therefore differ from the SM benchmark measurement from tree-level $B \to D^{(*)}K^{(*)}$ decays~\cite{LHCb-PAPER-2021-033,LHCb-CONF-2025-003}. 
Additionally, results on the branching fractions and \CP asymmetries of decay processes involving the $K^*(892)$ or $\rho(770)$ resonances can be used to cleanly test the SM through isospin sum rules, in the same way as for \decay{B}{K\pi} decays~\cite{Lipkin:1991st,Gronau:2005kz,Gronau:2010dd}.

This Letter contains a summary of results pertaining to \CP violation obtained in an amplitude analysis of \decay{\Bp}{\Kp\pip\pim} decays.
An extensive description of the experimental procedure along with a comprehensive collection of results can be found in a companion publication~\cite{BuToKpPipPimPRD}, while another companion Letter~\cite{BuToKpPipPimPRLStrong} details significant \CP-conserving implications arising from improved modelling of the hadronic structure that also facilitate identification of the underlying sources of \CP violation presented here. 
The analysis is based on \proton\proton collision data recorded with the LHCb detector in 2011--2012, corresponding to an integrated luminosity of 3\invfb. 
The \lhcb detector is a single-arm forward spectrometer covering the pseudorapidity range $2 < \eta < 5$, described in detail in Refs.~\cite{LHCb-DP-2008-001,LHCb-DP-2014-002}.

The selection of signal candidates follows procedures established for measurements of the relative branching fractions of inclusive \decay{\Bp}{h^+ h^{\prime+} h^{\prime-}} decays, where $h^{(\prime)}$ is a pion or kaon, based on the same data sample~\cite{LHCb-PAPER-2020-031}. 
Several minor adjustments to the selection requirements are made. 
As the \jpsi meson is not considered as signal in this analysis, its contribution is removed with a veto placed on the \pip\pim mass. 
The background from \decay{\Bp}{\Dzb h^+} decays is further suppressed with respect to Ref.~\cite{LHCb-PAPER-2020-031} with widened vetoes in the \Kp\pim and \pip\pim masses around the \Dzb peaks. 
Despite these, residual tails remain due to particle misidentification. They are suppressed by more stringent particle identification requirements in regions immediately surrounding the vetoed \Dzb mass regions.

After application of the selection criteria, the \Bp-candidate mass distribution is modelled to obtain signal and background yields. 
The fit function includes components for signal decays, combinatorial background, four-body partially reconstructed decays and misidentified three-body peaking backgrounds. 
Parameters describing the shapes of the signal and combinatorial background are determined directly from data, while those of the remaining backgrounds are fixed from dedicated simulation samples corrected for differences between data and simulation. 
The signal region in the \Bp-candidate mass, defined as three times the signal mass resolution on both sides of the \Bp peak, is retained for the subsequent amplitude analysis, comprising 115\,102 \decay{\Bp}{\Kp\pip\pim} candidates with an estimated purity of $(88.7\pm0.3)$\%.

Given the large number of broad overlapping resonances and coupled-channel thresholds, it is particularly challenging to obtain satisfactory phenomenological models for the \Kp\pim and \pip\pim S-wave amplitudes.
Therefore, three complementary approaches describing the total S-wave amplitude, $A_{\rm S}^\pm$ for \Bpm decays, are considered. 
In each case, a conventional ``isobaric'' paradigm for nonzero-spin resonances is used, in which the factorisation of two-body dynamics from three-body unitarity is the central assumption~\cite{isobar1,isobar2,isobar3}. 
The first model~(Isobar) applies the same approach to the S-wave, which comprises the coherent sum of directly produced intermediate scalar states. 
These include a generalised LASS~(GLASS) function~\cite{Estabrooks:1978de,Aston:1987ir,BaBar:2008inr} describing the \Kp\pim S-wave, together with $\pi\pi$--$\pi\pi$~\cite{Garcia-Martin:2011iqs} and $\pi\pi$--$K\Kbar$~\cite{Pelaez:2020gnd} scattering terms, both of which are 
multiplied by a form factor accounting for the production of the two-body system in the $B$-meson decay~\cite{AlvarengaNogueira:2015wpj,Bediaga:2013ela}. 
The second replaces the \pip\pim S-wave model with the K-matrix formalism, which enforces two-body unitarity with parameters obtained from scattering data~\cite{Aitchison:1972ay,Chung:1995dx,Anisovich:2002ij,BaBar:2008inr,FOCUS:2003tdy}.
The third implements a quasi-model-independent~(QMI) approach, marking the first instance in which any overlapping crossing partial waves in two different two-body systems produced in particle decay are simultaneously modelled; magnitudes and phases representing individual S-waves are free to vary in intervals~(bins) of the \Kp\pim and \pip\pim masses.

The signal amplitude for \Bp and that for \Bm signal decays is constructed as the sum over $N$ nonzero-spin resonant contributions and the aforementioned S-wave component, \mbox{$A^\pm(\Phi_3) \equiv \sum_{j=1}^N a_j^\pm A_j(\Phi_3) + A_{\rm S}^\pm(\Phi_3)$}, where $\Phi_3$ denotes the candidate position in three-body phase space. 
The $A_j$ terms describe the \CP-conserving dynamical amplitude of resonance $j$, consisting of: the product of a relativistic Breit--Wigner mass propagator or an appropriate variant~\cite{Bugg:2006gc,Flatte:1976xu,Pelaez:2022qby,GS}, the spin-dependent angular distribution~\cite{Zemach:1963bc,Zemach:1968zz} and Blatt--Weisskopf barrier factors~\cite{Blatt:1952ije,BlattWeisskopf}. 
All established intermediate resonant states~\cite{PDG2020} known to decay either into \Kp\pim or \pip\pim with poles below the \Dzb mass are included in this model. 
The complex coefficients $a_j^\pm \equiv (x_j \pm \Delta x_j) + i(y_j \pm \Delta y_j)$ parametrise the relative contributions of each resonance, where the positive~(negative) signs apply to \Bp~(\Bm) decays. 
The \CP-violating differences between \Bp and \Bm amplitudes are encoded in the freedom of $a_j^\pm$ and the relevant parameters of $A_{\rm S}^\pm$ to take different values for the charge-conjugate processes. 
As there is no sensitivity to the absolute phase of the \Bp and \Bm amplitudes, the coefficients of the largest non-S-wave contribution, the $K^*(892)^0$ resonance, relative to which coefficients of the remaining contributions are measured~(specifically, $x = 1$ and $y = \Delta y = 0$ are fixed for this component).

An event likelihood is constructed from the squared magnitude of the signal amplitude modulated by the efficiency variation across the Dalitz plot and convolved with the detector resolution in $m^2_{\pip\pim}$, while background model contributions are derived from data in the \Bp-mass sidebands or calibrated simulation. 
The Isobar and K-matrix approaches are implemented in the \textsc{Laura++} software package~\cite{Laura++}, while a GPU-accelerated fitter based on \textsc{Mint2}~\cite{Mint2} is used for the QMI approach. 

The quasi-two-body \CP-violation parameter for a component $j$, $\mathcal{A}^j_{\CP}$, is derived from an asymmetry between the \Bm and \Bp fit fractions, $\mathcal{F}^\pm_{j}$,
\begin{equation}
    \mathcal{A}^j_{\CP} \equiv \frac{\mathcal{F}^-_{j}-\mathcal{F}^+_{j}}{\mathcal{F}^-_{j}+\mathcal{F}^+_{j}}\,, 
    \quad\text{where}\quad
    \mathcal{F}^\pm_{j} \equiv \frac{\int |a^\pm_j A_j(\Phi_3)|^2 \textrm{d}\Phi_3}{\int \left(|A^+(\Phi_3)|^2 + |A^-(\Phi_3)|^2\right) \textrm{d}\Phi_3} \,.
    \label{eq:FFdef}
\end{equation}
Such \CP\ violation in resonant contributions can be driven purely by the Bander--Silverman--Soni mechanism~\cite{Bander:1979px}. 
For a single resonance, the strong-phase difference between two contributing amplitudes with different weak phases can also produce a nonzero \CP-violating phase difference between \Bm\ and \Bp\ couplings, $\arg(a_i^-) - \arg(a_i^+)$. 
Finally, the violation of \CP symmetry can also manifest between amplitudes of different spin as a difference between interference fractions for \Bm and \Bp decays, a long-distance phenomenon first identified in Ref.~\cite{AlvarengaNogueira:2015wpj} and observed in \decay{\Bp}{\pip\pip\pim} decays~\cite{LHCb-PAPER-2019-018,LHCb-PAPER-2019-017}. Interference fractions are defined as
\begin{equation}
    \mathcal{I}^\pm_{j\,<\,k} \equiv \frac{\int 2\Real[a^\pm_j a^{\pm *}_k A_j(\Phi_3) A^*_k(\Phi_3)] \textrm{d}\Phi_3}{\int \left(|A^+(\Phi_3)|^2 + |A^-(\Phi_3)|^2\right) \textrm{d}\Phi_3} \,.
    \label{eq:IFFdef}
\end{equation}
Thus, any well-defined measure of such a difference will need to take into account that interference fractions are signed quantities. 
This is achieved through the combined fit fraction for the coherent sum of two components, $\mathcal{F}^\pm_{jk}$, and its asymmetry, $\mathcal{A}^{jk}_{\CP}$, which are defined in the same way as Eq.~\eqref{eq:FFdef}.
Then a simple expansion gives 
\begin{equation}
    \mathcal{A}^{jk}_{\CP} \equiv 
    \mathcal{A}^j_{\CP}\frac{\mathcal{F}_j}{\mathcal{F}_{jk}} + 
    \mathcal{A}^k_{\CP}\frac{\mathcal{F}_k}{\mathcal{F}_{jk}} +
    2\mathcal{A}^{jk}_{\mathcal{I}}\,, 
    \quad\text{with}\quad \mathcal{A}^{jk}_{\mathcal{I}} \equiv \frac{\mathcal{I}^-_{jk} - \mathcal{I}^+_{jk}}{\mathcal{F}_{jk}} \,,
    \label{eq:ACP-IFFdef}
\end{equation}
where $\mathcal{F}_{j(k)} = \mathcal{F}^-_{j(k)}+\mathcal{F}^+_{j(k)}$.
The first two terms re-express the quasi-two-body \CP asymmetry in their respective components, while the third term codifies the \CP asymmetry in their interference.

The main sources of experimental systematic uncertainty, detailed in Ref.~\cite{BuToKpPipPimPRD}, originate from: the \B-candidate mass model from which fixed signal and background yields are propagated; the inherent uncertainty in calibration as it pertains to efficiency and background phase-space distributions from simulation further weighted by experimentally determined amplitude models~\cite{LHCb-PAPER-2019-017,LHCb-PAPER-2019-018,LHCb-PAPER-2018-051,BESIII:2017kyd}; and definition of the sideband region in data from which the combinatorial background distribution is taken. 
Systematic uncertainties related to the physical amplitude models are also considered. 
These comprise the variation of fixed masses and widths of established resonances according to their known uncertainties~\cite{PDG2020} and alternative barrier factor radii, in addition to the inclusion of the $K^*_4(2045)^0$ and more speculative resonant structures such as the $f_2(1430)$.

Systematic uncertainties on the reported \CP-asymmetry parameters are typically comparable to the statistical uncertainties.
This indicates a good level of control over the models and experimental effects such as backgrounds, motivating future studies with more data. 
The K-matrix systematic uncertainties tend to be smaller compared to the other S-wave approaches, which is a consequence of the scattering parameters taken from Ref.~\cite{Anisovich:2002ij} not being published with uncertainties.
The systematic uncertainties in the QMI approach tend to be largest, due to the assumption that the S-wave amplitudes are constant within each bin. 

Selected results for \CP-asymmetry parameters are given in Table~\ref{tab:acp}, with the remainder reported in the Appendix. 
For the leading resonant contributions in the \Kp\pim system, the $K^*(892)^0$ and $K^*_0(1430)^0$~(extraction of $K^*_0(1430)^0$ information from the GLASS function is discussed in a companion article~\cite{BuToKpPipPimPRD}), the precision on these measurements sets the most stringent limits on the only other potential contribution to these modes beyond a single gluonic-penguin amplitude: a penguin-annihilation amplitude~\cite{Gronau:1994rj}. This is of great importance in clean tests of the SM through \decay{B}{K\pi} isospin sum rules for excited intermediate states~\cite{Lipkin:1991st,Gronau:2005kz,Gronau:2010dd}.

\begin{table}[!tb]
    \centering
    \caption{\label{tab:acp} 
        Selected \CP-violation results for the three S-wave approaches. 
        For all measurements, the first uncertainty is statistical while the second is systematic. 
        The significance~($\sigma$) of each deviation from conservation of \CP symmetry is determined using the total uncertainty. 
    }

    \resizebox{\linewidth}{!}{
    \renewcommand{\arraystretch}{1.1}
    \begin{tabular}
    {@{\hspace{0.0cm}}l@{\hspace{0.0cm}}
     @{\hspace{0.125cm}}c@{\hspace{0.0cm}}
     @{\hspace{0.125cm}}r@{\hspace{0.125cm}}
     @{\hspace{0.125cm}}c@{\hspace{0.0cm}}
     @{\hspace{0.125cm}}r@{\hspace{0.125cm}}
     @{\hspace{0.125cm}}c@{\hspace{0.0cm}}
     @{\hspace{0.125cm}}r@{\hspace{0.0cm}}} \hline
    Parameter & \phantom{--}Isobar & & \phantom{--}K-matrix & & \phantom{--}QMI & \\ \hline

    $\mathcal{A}_{\CP}(K^*(892)^0)$ & $-0.038 \pm 0.010 \pm 0.018$ & $(1.9\sigma)$ & $-0.035 \pm 0.012 \pm 0.017$ & $(1.7\sigma)$ & $-0.027 \pm 0.014 \pm 0.023$ & $(1.0\sigma)$\\

    $\mathcal{A}_{\CP}(K^*_0(1430)^0)$ & $-0.094 \pm 0.020 \pm 0.046$ & $(1.9\sigma)$ & $-0.121 \pm 0.015 \pm 0.050$ & $(2.3\sigma)$ & ---& \\\hline

    $\mathcal{A}_{\CP}(\rho(770)^0)$ & $+0.280 \pm 0.022 \pm 0.029$ & $(7.6\sigma)$ & $+0.271 \pm 0.022 \pm 0.015$ & $(>\!10\sigma)$ & $+0.359 \pm 0.037 \pm 0.048$ & $(5.9\sigma)$\\

    $\mathcal{A}^{\rm S}_{\CP}(m_{\pip\pim} < 4m_{\pi})$ & $+0.665 \pm 0.032 \pm 0.053$ & $(>\!10\sigma)$ & $+0.638 \pm 0.008 \pm 0.016$ & $(>\!10\sigma)$ & $+0.754 \pm 0.115 \pm 0.113$ & $(4.7\sigma)$\\

    $\mathcal{A}^{\rm SD}_{\mathcal{I}}$ $(\times 10^{-1})$ & $+0.058 \pm 0.007 \pm 0.008$ & $(5.2\sigma)$ & $+0.055 \pm 0.007 \pm 0.007$ & $(5.3\sigma)$ & $+0.071 \pm 0.014 \pm 0.009$ & $(4.4\sigma)$\\

    $\mathcal{A}^{\rm SP}_{\mathcal{I}}$ $(\times 10^{-1})$ & $-0.237 \pm 0.018 \pm 0.040$ & $(5.4\sigma)$ & $-0.249 \pm 0.021 \pm 0.040$ & $(5.5\sigma)$ & $-0.253 \pm 0.025 \pm 0.028$ & $(6.7\sigma)$\\

    $\mathcal{A}_{\CP}(\rho_3(1690)^0)$ & $+0.603 \pm 0.090 \pm 0.121$ & $(4.0\sigma)$ & $+0.477 \pm 0.100 \pm 0.096$ & $(3.4\sigma)$ & $+0.698 \pm 0.135 \pm 0.163$ & $(3.3\sigma)$\\

    \hline
    \end{tabular}
    }
\end{table}

\begin{figure}[!t]
    \centering
    \includegraphics[width=0.5\textwidth]{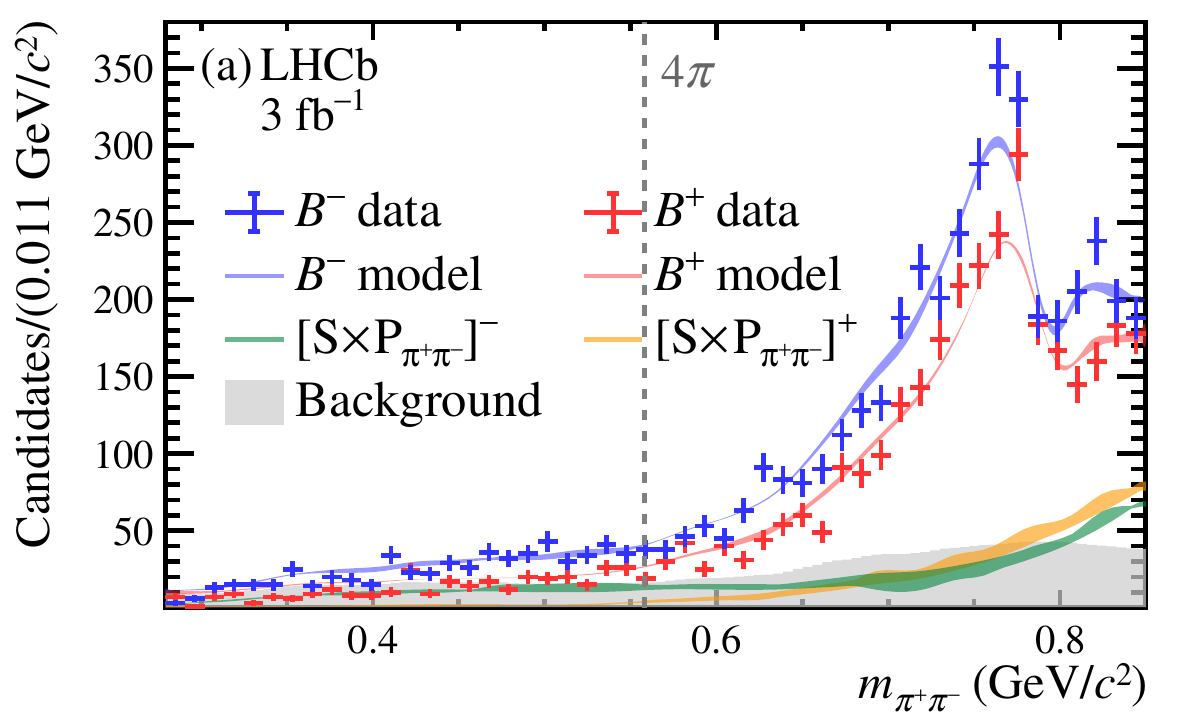}%
    \includegraphics[width=0.5\textwidth]{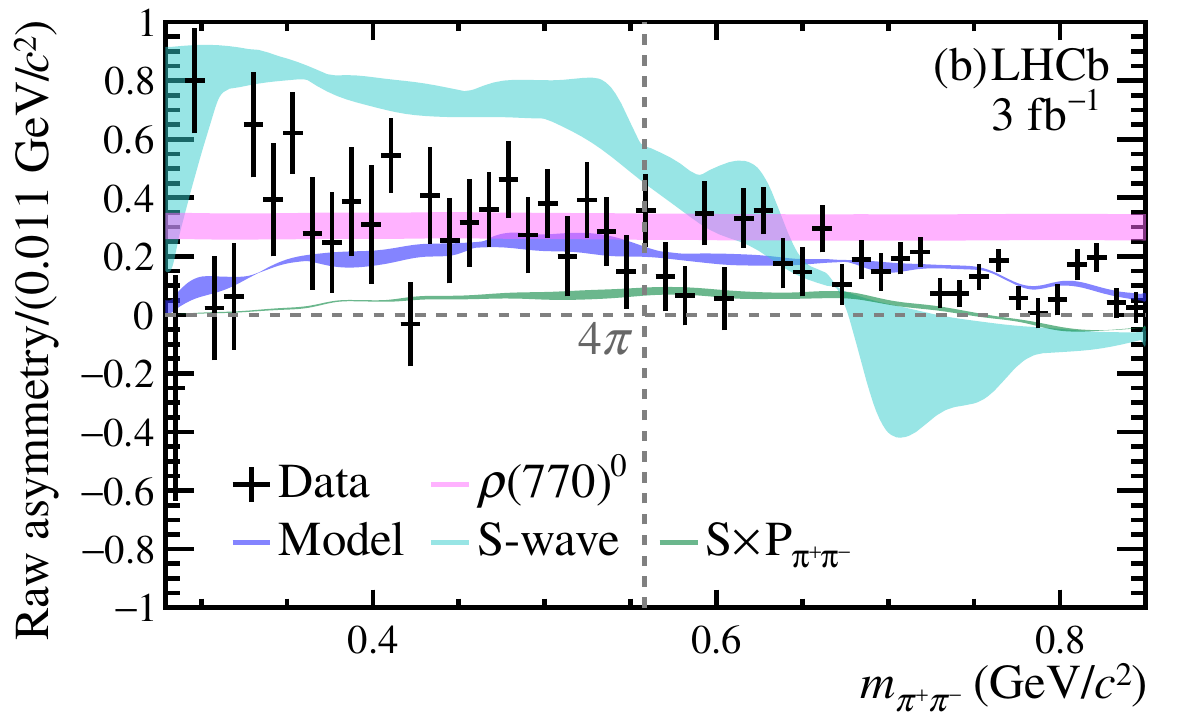}

    \caption{
    Projections of the data on the $\pi^+ \pi^-$ mass from threshold  up to $0.85\gevcc$, with results of the fits superimposed, for (a)~$B^-$ and $B^+$ decays separately and (b)~the raw asymmetry between them. 
    The width of the model curves indicate the full spread in central values obtained with the different S-wave approaches, and the background contribution is $C\!P$ averaged.
    Different components of the model are also shown, as indicated in the legends, with interference contributions between two components denoted by the $\times$ symbol. 
    The vertical dashed lines indicate the kinematic threshold of the $4\pi$ rescattering channel. 
    }
    \label{fig:rho}
\end{figure}

Projections of the data in the low $\pip\pim$ mass region from threshold up to \mbox{$m_{\pip\pim} = 0.85\gevcc$} are shown in Fig.~\ref{fig:rho}, and compared to the fitted amplitude models.
The spread between the results of fits with different S-wave approaches is shown by the width of the model curve. 
The $\rho(770)^0$ resonance is prominent in this mass projection, and the asymmetry between its contribution in \Bm and \Bp data is clearly visible.
The measured value of its quasi-two-body \CP-violation parameter, given in Table~\ref{tab:acp}, is in agreement with theory expectations~\cite{Beneke:2003zv,Cheng:2009cn,Cheng:2016shb,Li:2006jv,Chai:2022ptk,Cheng:2014rfa,Bell:2015koa}, and has a minimum significance among S-wave models of 5.9~standard deviations~$(\sigma)$.
This constitutes the first observation of quasi-two-body \CP violation in \decay{\Bp}{\rho(770)^0\Kp} decays. 
Further to this, evidence for \CP violation mediated by final-state interactions through a nonzero phase difference is also noteworthy at a minimum significance of $3.6\sigma$, as estimated from Table~\ref{tab:end:dphase} with the $z$-score.

Another striking feature of the data in the low $m_{\pip\pim}$ region is the effect of $\rho^0$--$\omega$ mixing. 
Contrary to some theoretical predictions~\cite{Li:2006jv,Wang:2008rk,Cheng:2009cn,Chai:2022ptk}, no evidence for \CP-violation associated with this effect is found.
This explains the apparent discrepancy between the $\mathcal{A}_{\CP}(\rho(770)^0)$ values reported in Table~\ref{tab:acp} and the measurement of the model-independent P-wave asymmetry in the $\rho(770)^0$ region, $0.150 \pm 0.022$~\cite{LHCb-PAPER-2021-050}: the absence of significant \CP violation in the sizeable $\rho^0$--$\omega$ mixing contribution dilutes the overall \CP asymmetry in the \pip\pim P-wave in this region. 

At very low \pip\pim mass, below the four-pion~($4\pi$) kinematic threshold and thus barely affected by the $\rho(770)^0$ tail and \Kp\pim reflections, a large positive raw asymmetry is observed. 
All three S-wave approaches give consistent descriptions of this effect, which match the data well down to $m_{\pip\pim} \sim 0.4 \gevcc$ but cannot describe the asymmetry observed immediately above the \pip\pim threshold. 
This may indicate that the strength of \cquark\cquarkbar loop contributions in the \Kp\pim partial wave is considerable~\cite{El-Bennich:2009gqk,Heuser:2025mnk}. 
Nonresonant isospin-2 contributions may also play an essential role~\cite{Heuser:2026glv}. 
The significance of the asymmetry in the inclusive \pip\pim S-wave beneath the $4\pi$ kinematic threshold, $\mathcal{A}^{\rm S}_{\CP}$, is found to be $4.7\sigma$ with the QMI approach, with larger significance values in the Isobar and K-matrix approaches. 
This disparity is due to flexibility of the binned QMI S-wave model in this region over the other approaches whose lineshapes are fixed across the full range of $m_{\pip\pim}$.

\begin{figure}[!tb]
    \centering
    \includegraphics[width=0.5\textwidth]{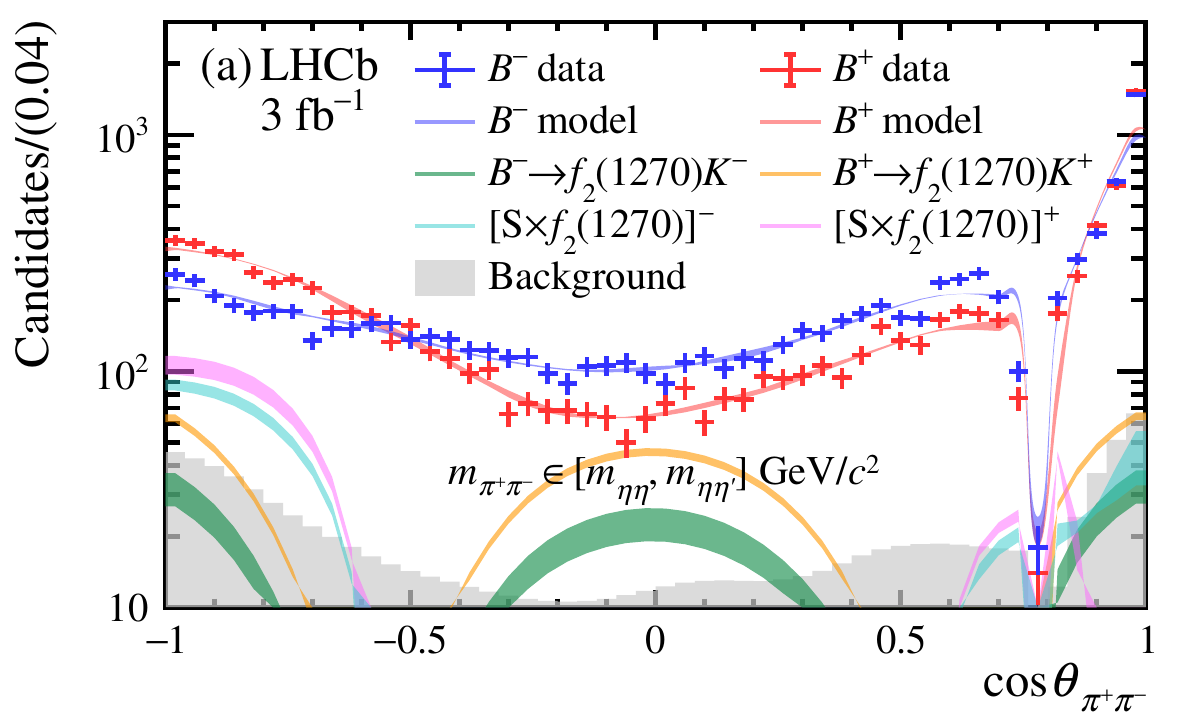}%
    \includegraphics[width=0.5\textwidth]{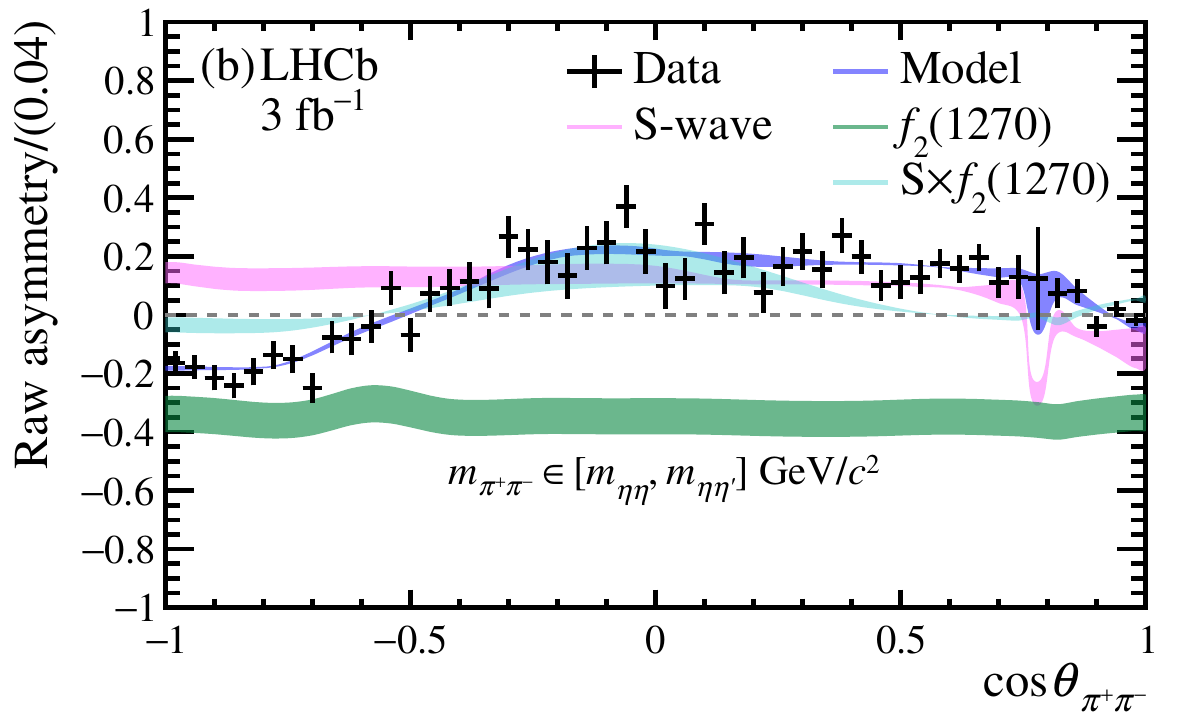}

    \includegraphics[width=0.5\textwidth]{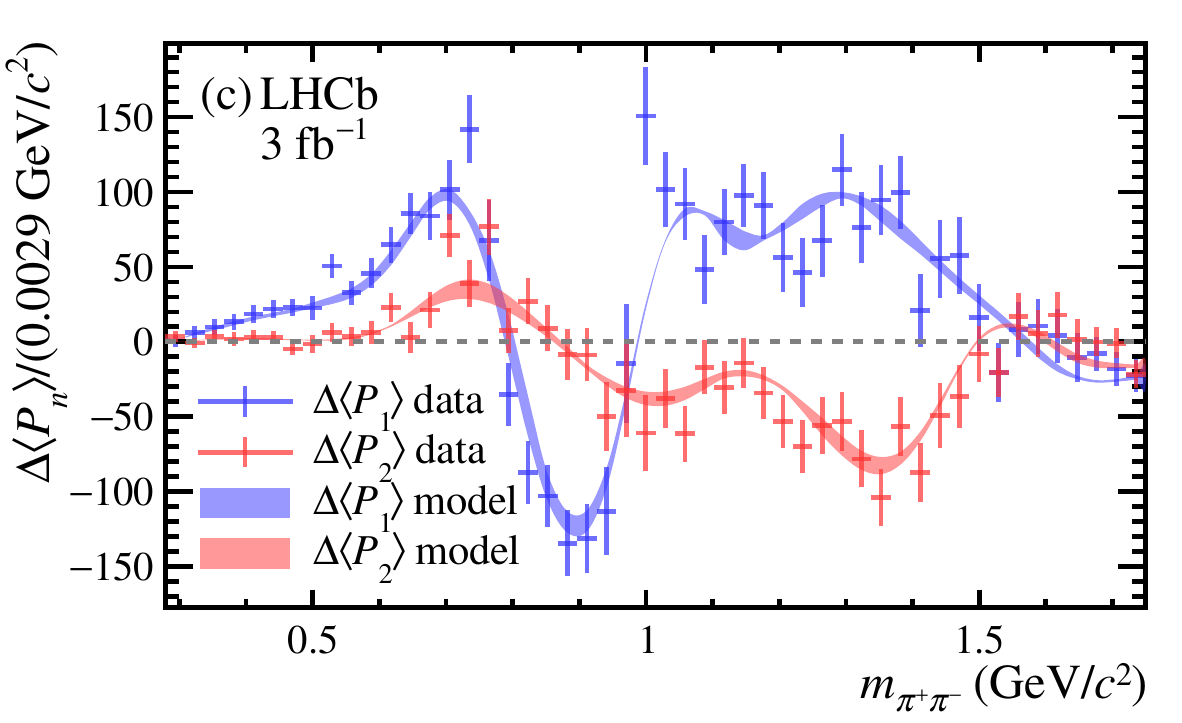}%
    \includegraphics[width=0.5\textwidth]{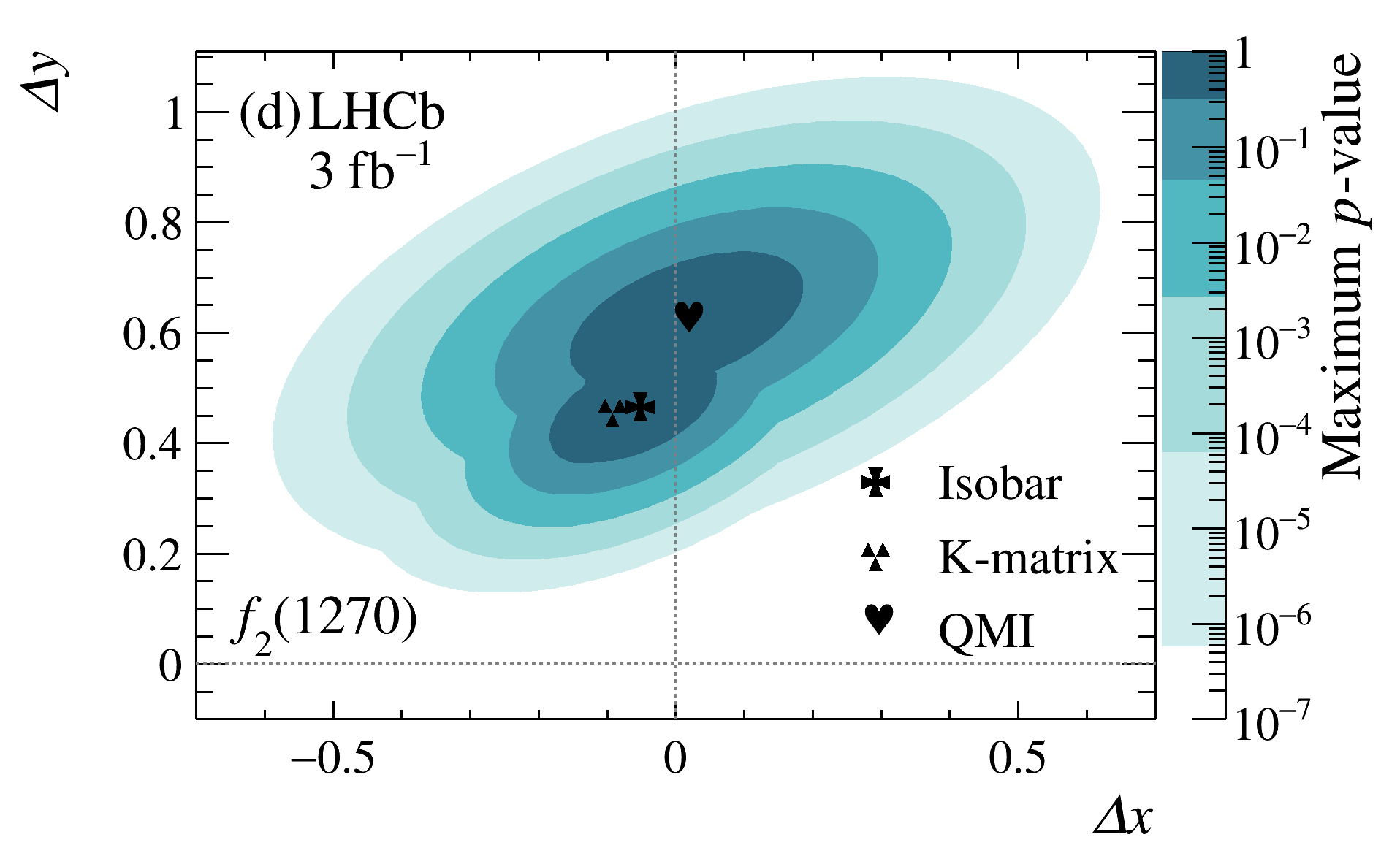}

    \caption{Projections of the data, and comparison with the fit results, onto (a)~the $\cos\theta_{\pi^+ \pi^-}$ distributions from $B^-$ and $B^+$ decays above a $C\!P$-averaged background contribution, (b)~their corresponding $C\!P$ asymmetry in the $f_2(1270)^0$ region, and (c)~the Legendre moment-weighted differences between $B^-$ and $B^+$ in the charmless $\pi^+ \pi^-$ region. The widths of the model curves indicate the full spread in central values obtained with the different S-wave approaches, and 
    interference contributions between two components are denoted by the $\times$ symbol in the relevant legends. 
    Finally, (d)~illustrates contours of the maximum $p$-value among S-wave approaches corresponding to the total uncertainty profile of the $C\!P$-violating amplitude coefficients for the $f_2(1270)$ contribution.}
    \label{fig:f2}
\end{figure}

In the region between the $\eta\eta$ and $\eta\eta^\prime$ thresholds corresponding to the range ${m_{\pip\pim} \in [1.10, 1.51] \gevcc}$, the leading contributions arise from the $f_2(1270)$ resonance and the \pip\pim S-wave, although reflections from \Kp\pim contributions are also significant.
The data and fit models in this region, projected onto the $\cos\theta_{\pip\pim}$ variable, are shown in Figs.~\ref{fig:f2}(a) and~(b). 
Here, $\theta_{\pip\pim}$ is the \pip\pim helicity angle, defined as the angle between the momentum of the pion with opposite charge to the $B$ meson and that of the kaon from the $B$ decay, evaluated in the \pip\pim rest frame. 
The effect of \CP violation mediated by the $f_2(1270)$ resonance is most clearly seen in the range $\cos\theta_{\pip\pim} < -0.5$, where the sign of the asymmetry in data is consistent.
The raw asymmetry in this region is diluted by the \pip\pim S-wave, which has an opposite asymmetry, with this term dominating at $\cos\theta_{\pip\pim}$ values near zero.
For $\cos\theta_{\pip\pim}$ values above the $\Dzb \to \Kp \pim$ veto (apparent near $+0.8$), reflections from \Kp\pim contributions become dominant, leading to negligible overall asymmetry.

Effects due to interference between the $f_2(1270)$ resonance and the \pip\pim S-wave are also visible in Figs.~\ref{fig:f2}(a) and~(b).
This contribution is positive at the extremities of the $\cos\theta_{\pip\pim}$ distribution, where it has a small negative asymmetry, and is negative in the central region, where its asymmetry therefore has the opposite sign.
The negative contributions, which arise due to destructive interference, are not visible in Fig.~\ref{fig:f2}(a). 
The amount of \CP violation in S-D interference is quantified through $\mathcal{A}^{\rm SD}_{\mathcal{I}}$, and found to have a significance of at least $4.4\sigma$.

Weighting data by Legendre polynomials of the cosine of helicity angle is a common procedure in amplitude analyses used to visualise contributions from different partial waves and their interferences.  
By weighting the \Bm and \Bp samples separately, and then taking the difference, this approach can also be used to inspect corresponding asymmetries. 
The differences in the data weighted with first- and second-order Legendre polynomials, denoted $\Delta\left<P_1\right>$ and $\Delta\left<P_2\right>$ respectively, are shown in Fig.~\ref{fig:f2}(c), as a function of $m_{\pip\pim}$ up to the open-charm threshold.
The former has a large contribution from interference between S- and P-waves, and is seen to change sign at both the $\rho(770)^0$ and $f_0(980)$ poles, a similar behaviour to that observed in \decay{\Bp}{\pip\pip\pim} decays~\cite{LHCb-PAPER-2019-018,LHCb-PAPER-2019-017}.
These sign flips are associated with known rapid phase variations at those masses of the P- and S-waves, respectively.
The significance of \CP violation in interference between the S-wave and $\rho(770)^0$ specifically is confirmed at $5.4\sigma$, denoted as $\mathcal{A}^{\rm SP}_{\mathcal{I}}$ in Table~\ref{tab:acp}.
The $\Delta\left<P_2\right>$ distribution, on the other hand, includes contributions from both the P-wave (with the asymmetry due to the $\rho(770)^0$ contribution clearly visible) and interference between S- and D-waves.
The latter is responsible for the structure in the $f_2(1270)$ region, as discussed above. 

The quasi-two-body \CP-violation parameter of the $f_2(1270)$ contribution agrees with theoretical predictions~\cite{Cheng:2010yd,Li:2018lbd}, though does not reach the significance threshold for observation of quasi-two-body \CP violation in all S-wave approaches. 
As a search for \CP violation through the phase difference is similarly inconclusive, the significance of \CP violation is assessed at amplitude level. 
A $\chi^2$ test-statistic is constructed from the central values and total covariance matrix of the raw fit result, and a $p$-value scan~\cite{CKMfitter2005} of the \CP-violating couplings of the $f_2(1270)$ contribution is performed, with results shown in Fig.~\ref{fig:f2}(d).
This procedure accounts for the impact of the remainder of the amplitude model. 
For two degrees of freedom, confidence intervals can be derived by which \CP conservation is rejected at the 8.7$\sigma$, 9.1$\sigma$ and 8.2$\sigma$ levels for the Isobar, K-matrix and QMI approaches, respectively. 

\begin{figure}[!tb]
    \centering
    \includegraphics[width=0.5\textwidth]{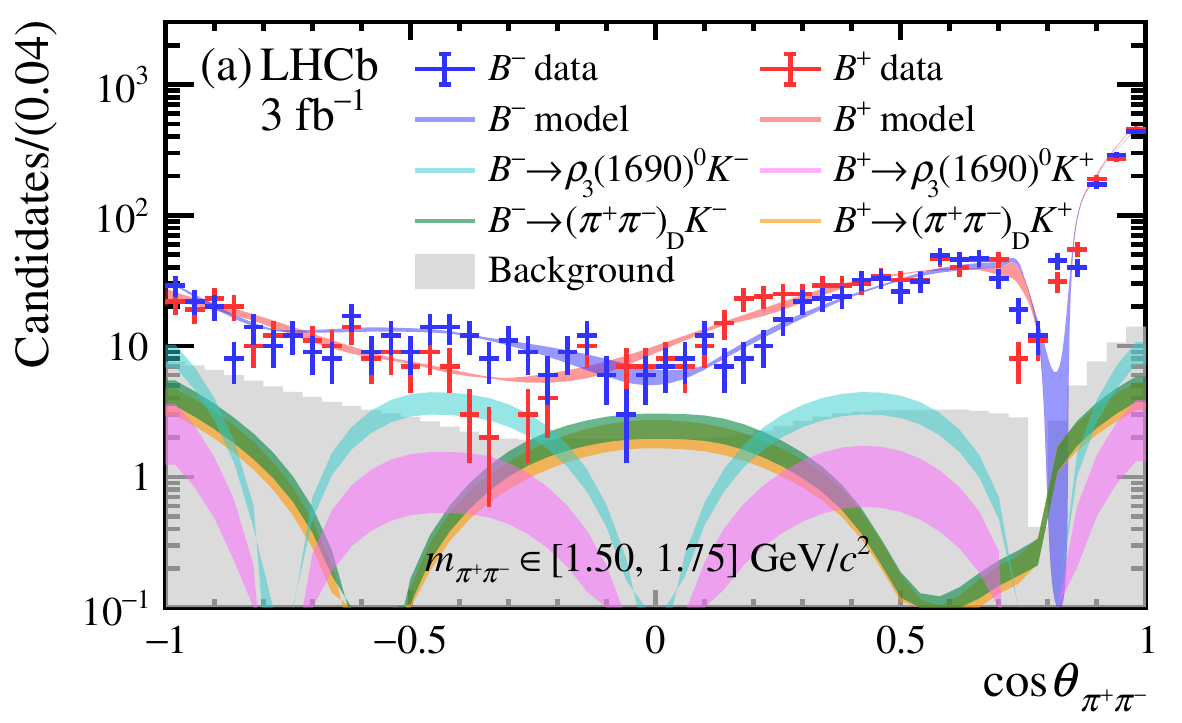}%
    \includegraphics[width=0.5\textwidth]{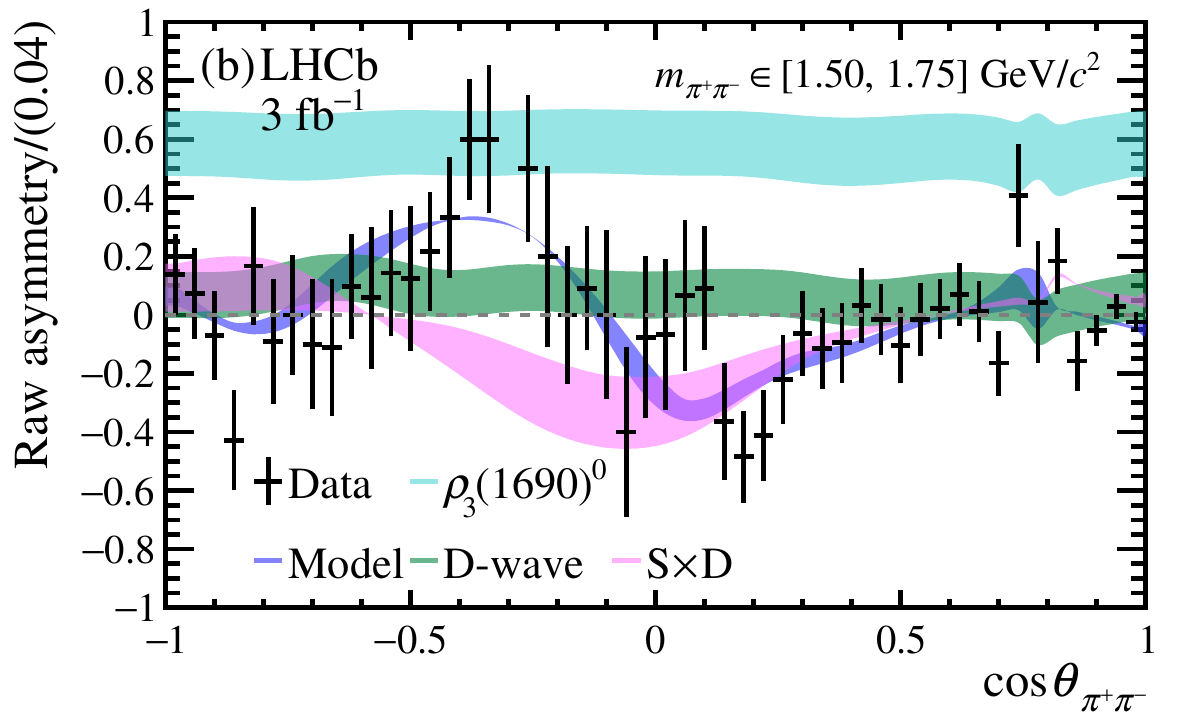}

    \caption{Model projections onto (a)~the $\cos\theta_{\pi^+ \pi^-}$ distributions from $B^-$ and $B^+$ decays atop a $C\!P$-averaged background and (b)~their corresponding $C\!P$ asymmetry in the $\rho_3(1690)^0$ region. The model uncertainties indicate the full spread in central values obtained for each S-wave approach.}
    \label{fig:rho3}
\end{figure}

Finally, Fig.~\ref{fig:rho3} shows the data and fit models projected onto $\cos\theta_{\pip\pim}$ in the region from $m_{\pip\pim} = 1.5\gevcc$ up to the open-charm threshold. 
The $\rho_3(1690)^0$ resonance contributes in this region, together with several other components. 
A large quasi-two-body \CP violation for the $\rho_3(1690)^0$ component is predominantly responsible for the asymmetry in data visible in the region $\cos\theta_{\pip\pim} \sim -0.5$.
In other regions, effects due to the \pip\pim D-wave, due to interference between S- and D-waves, and due to \Kp\pim reflections are important.
From Table~\ref{tab:acp}, the minimum significance of $\mathcal{A}_{\CP}(\rho_3(1690)^0)$ being nonzero is found to be $3.3\sigma$.

To conclude, an amplitude analysis of the \decay{\Bp}{\Kp\pip\pim} decay is performed using \proton\proton collision data recorded with the LHCb detector in 2011--2012, corresponding to an integrated luminosity of 3\invfb.
Three complementary approaches are used to describe the large S-wave contribution to this decay. 
Good agreement is found between all three models and the data. 
First observations of \CP violation in \decay{\Bp}{\rho(770)^0 \Kp} and \decay{\Bp}{f_2(1270) \Kp} decays, and in interference between \pip\pim S- and P-waves recoiling against a kaon, are reported. 
Evidence for \CP violation in \decay{\Bp}{\rho_3(1690)^0\Kp} is found, which is the first in any decay involving a spin-3 resonance.
The first evidence is also reported for \CP violation in the fully elastic $\pi\pi$--$\pi\pi$ S-wave rescattering region and in interference between S- and D-waves involving any decay process. 
As such, these results provide significant new insight into how \CP violation manifests in multibody $b$-hadron decays, and motivate further study into the processes that govern \CP violation in the charmless dipion mass dominated by $b \to s$ transitions.

%% file: LHCb/acknowledgements.tex
\section*{Acknowledgements}
%
%
\noindent We express our gratitude to our colleagues in the CERN
accelerator departments for the excellent performance of the LHC. We
thank the technical and administrative staff at the LHCb
institutes.
We acknowledge support from CERN and from the national agencies:
ARC (Australia);
CAPES, CNPq, FAPERJ and FINEP (Brazil); 
MOST and NSFC (China); 
CNRS/IN2P3 and CEA (France);  
BMFTR, DFG and MPG (Germany);
NKFIH (Hungary);              
INFN (Italy); 
NWO (Netherlands); 
MNiSW and NCN (Poland); 
MEC/IFA (Romania); 
MICIU and AEI (Spain);
SNSF and SER (Switzerland); 
NASU (Ukraine); 
STFC (United Kingdom); 
DOE NP and NSF (USA).
We acknowledge the computing resources that are provided by ARDC (Australia), 
CBPF (Brazil),
CERN, 
IHEP and LZU (China),
IN2P3 (France), 
KIT and DESY (Germany), 
INFN (Italy), 
SURF (Netherlands),
Polish WLCG (Poland),
IFIN-HH (Romania), 
PIC (Spain), CSCS (Switzerland), 
GridPP (United Kingdom),
and NSF (USA).  
We are indebted to the communities behind the multiple open-source
software packages on which we depend.
Individual groups or members have received support from
RTP (Australia), 
FWO Odysseus grant G0ASD25N (Belgium), 
Key Research Program of Frontier Sciences of CAS, CAS PIFI, CAS CCEPP (China); 
Minciencias (Colombia);
EPLANET, Marie Sk\l{}odowska-Curie Actions, ERC and NextGenerationEU (European Union);
A*MIDEX, ANR, IPhU and Labex P2IO, and R\'{e}gion Auvergne-Rh\^{o}ne-Alpes (France);
Alexander-von-Humboldt Foundation (Germany);
ICSC (Italy); 
Severo Ochoa and Mar\'ia de Maeztu Units of Excellence, GVA, XuntaGal, GENCAT, InTalent-Inditex and Prog.~Atracci\'on Talento CM (Spain);
the Leverhulme Trust, the Royal Society and UKRI (United Kingdom).

%% file: appendix.tex
\section*{End matter}
\label{sec:appendix}

\appendix

\section[Quasi-two-body CP-violation parameters and strong-phase differences]{\boldmath Quasi-two-body \CP-violation parameters and strong-phase differences}

\begin{figure}[t]
    \centering
    \includegraphics[width=0.5\linewidth]{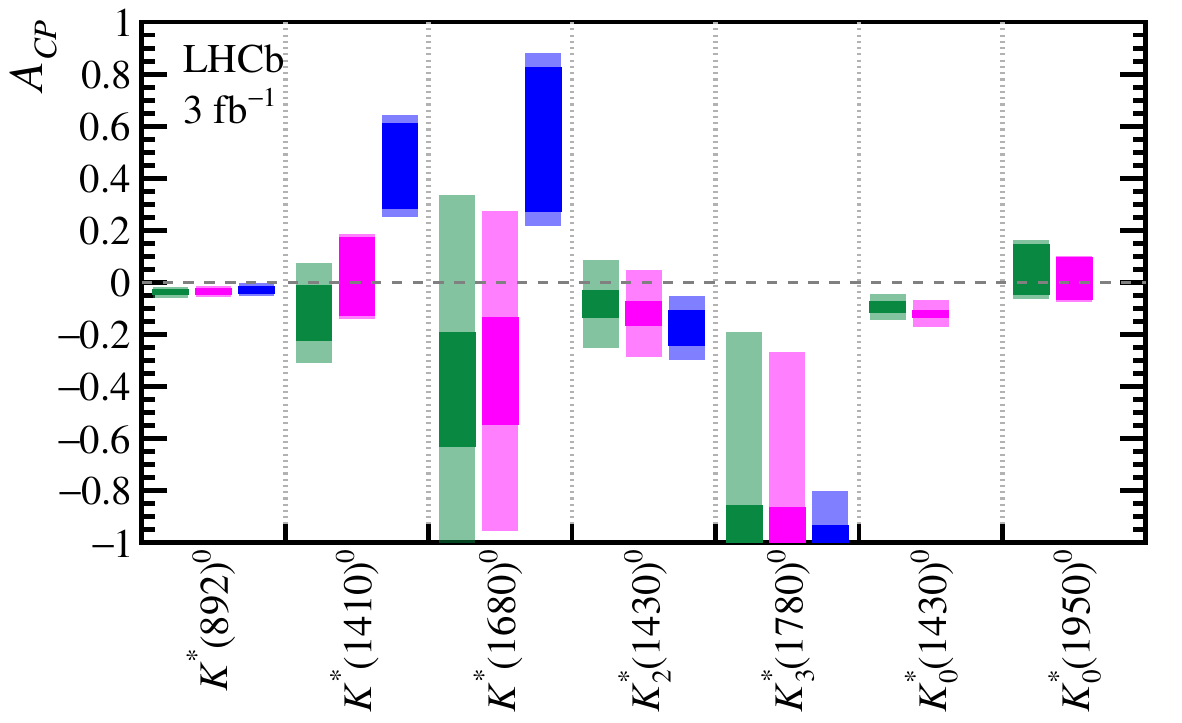}%
    \includegraphics[width=0.5\linewidth]{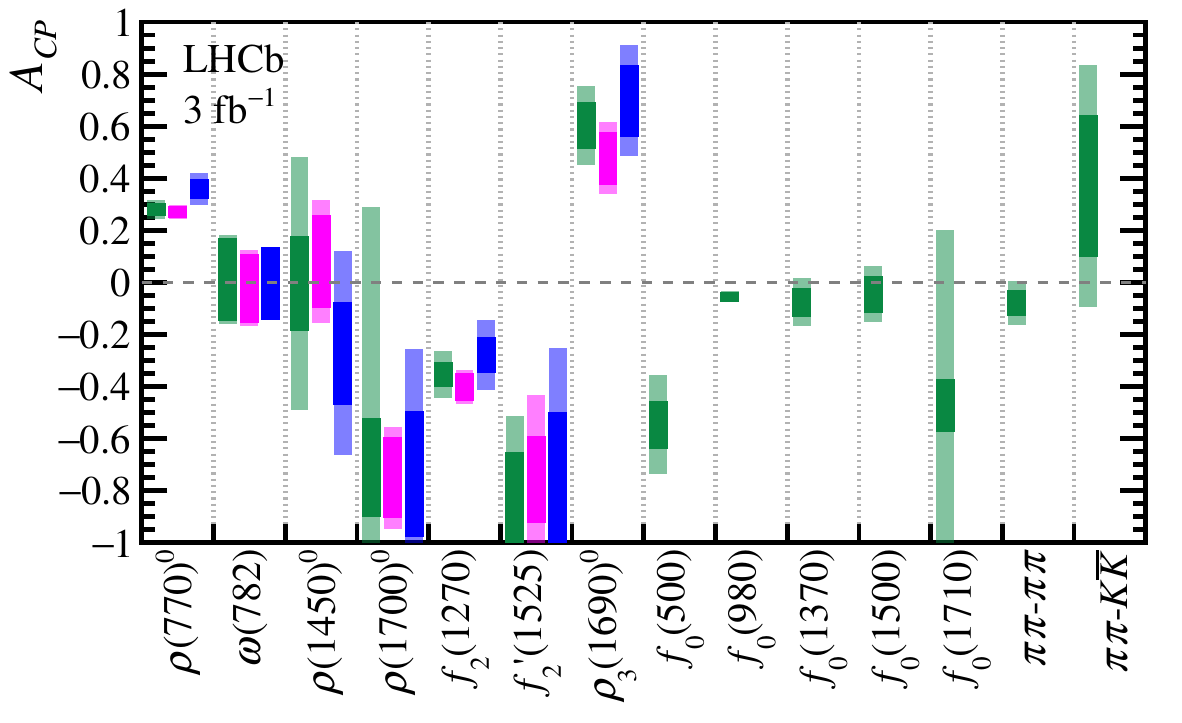}

    \caption{
        Comparison of the quasi-two-body asymmetries between Isobar~(green), K-matrix~(magenta) and QMI~(blue) for states in (left)~$K^+ \pi^-$ and (right)~$\pi^+ \pi^-$. Solid bands show the statistical spread while the lighter shading indicates ranges for the total uncertainty.
    }
    \label{fig:end:acp}
\end{figure}

\begin{figure}[tb]
    \centering
    \includegraphics[width=0.5\linewidth]{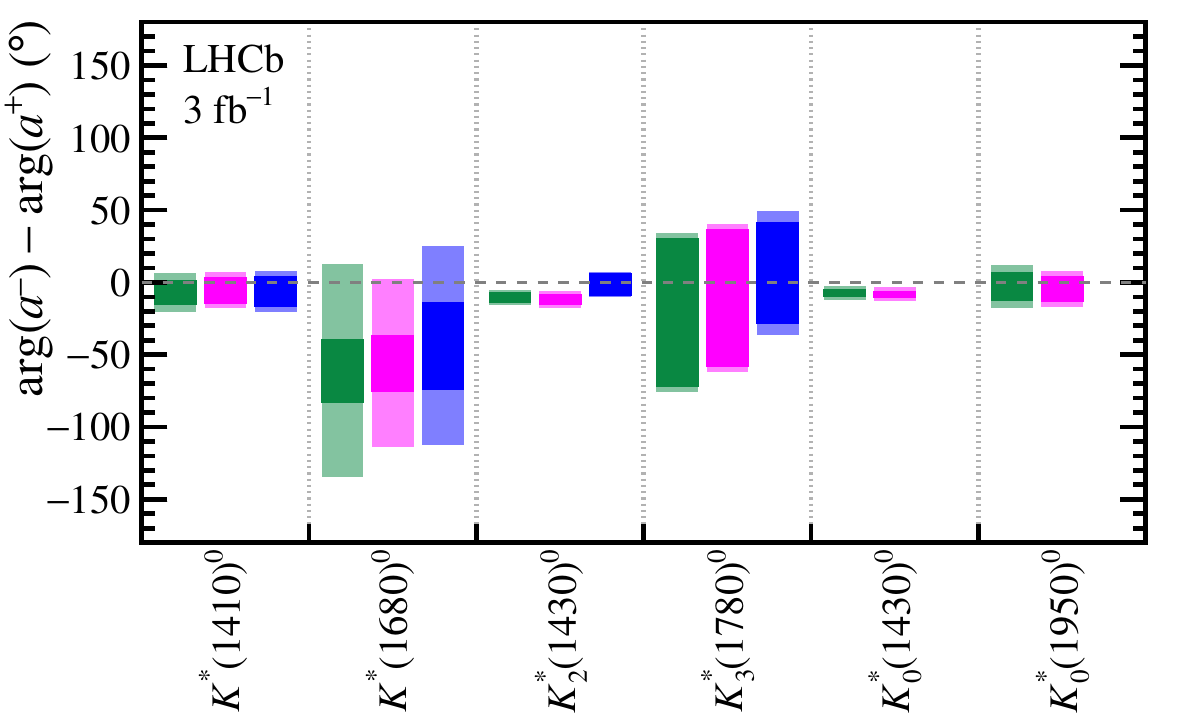}%
    \includegraphics[width=0.5\linewidth]{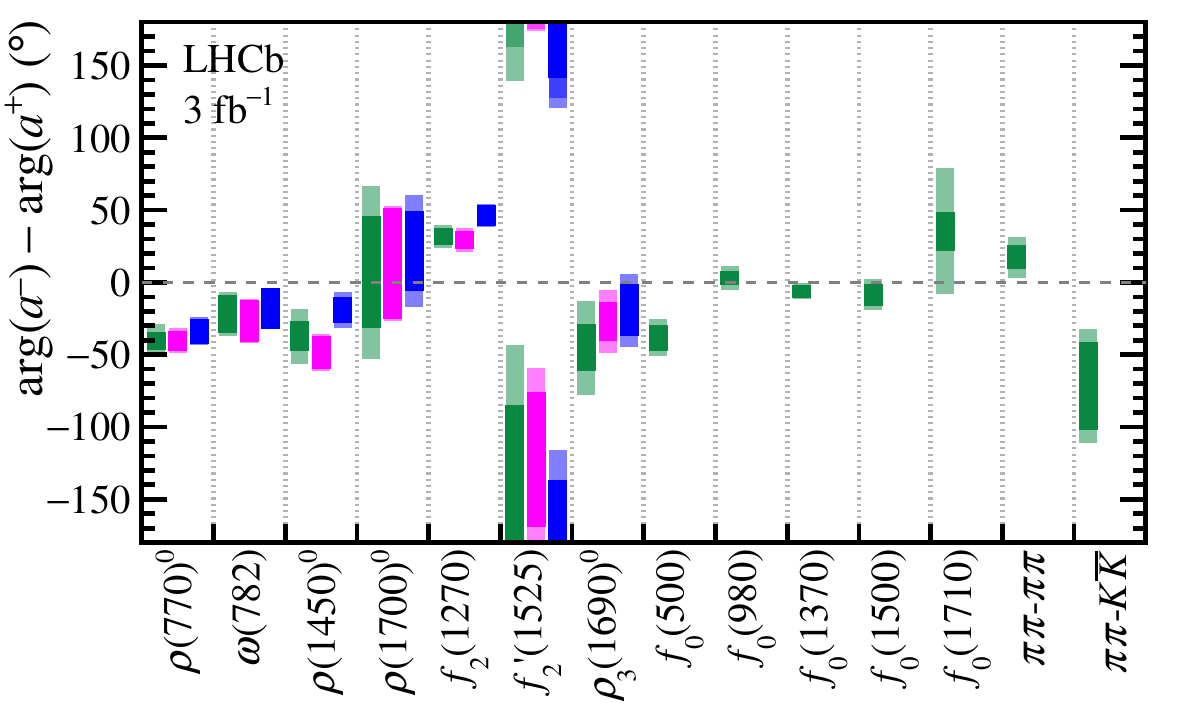}

    \caption{
        Comparison of the phase differences between Isobar~(green), K-matrix~(magenta) and QMI~(blue) for states in (left)~$K^+ \pi^-$ and (right)~$\pi^+ \pi^-$. Solid bands show the statistical spread while the lighter shading indicates ranges for  the total uncertainty.
    }
    \label{fig:end:dphase}
\end{figure}

For completeness, all \CP-violating observables that characterise the amplitude models are presented. The quasi-two-body \CP asymmetries for the remaining contributions not already included in Table~\ref{tab:acp} are reported in Table~\ref{tab:end:acp}, where to facilitate a meaningful comparison between S-wave approaches, the total \CP asymmetry is determined from the combined S-wave in \Kp\pim and \pip\pim. 
To ease comparison, all quasi-two-body \CP asymmetries are visually represented in Fig.~\ref{fig:end:acp}. 
The \CP-violating phase differences for each component are listed in Table~\ref{tab:end:dphase} and shown in Fig.~\ref{fig:end:dphase}.

\begin{table}[b]
    \centering
    \caption{\label{tab:end:acp}
        Remaining quasi-two-body \CP-violation results not already included in the main text for the three S-wave approaches.
        For all measurements, the first uncertainty is statistical while the second is systematic. 
    }
    \renewcommand{\arraystretch}{1.1}
    \begin{tabular}
    {@{\hspace{0.0cm}}l@{\hspace{0.125cm}}
     @{\hspace{0.125cm}}c@{\hspace{0.125cm}}
     @{\hspace{0.125cm}}c@{\hspace{0.125cm}}
     @{\hspace{0.125cm}}c@{\hspace{0.0cm}}} \hline
    Component & Isobar & K-matrix & QMI \\ \hline

$K^*(1410)^0$ & $-0.116 \pm 0.106 \pm 0.160$ & $+0.024 \pm 0.150 \pm 0.063$ & $+0.447 \pm 0.163 \pm 0.111$\\
$K^*(1680)^0$ & $-0.412 \pm 0.220 \pm 0.716$ & $-0.340 \pm 0.206 \pm 0.581$ & $+0.549 \pm 0.277 \pm 0.182$\\
$\omega(782)$ & $+0.011 \pm 0.158 \pm 0.064$ & $-0.023 \pm 0.131 \pm 0.062$ & $-0.004 \pm 0.139 \pm 0.015$\\
$\rho(1450)^0$ & $-0.005 \pm 0.180 \pm 0.452$ & $+0.081 \pm 0.178 \pm 0.156$ & $-0.273 \pm 0.197 \pm 0.339$\\
$\rho(1700)^0$ & $-0.711 \pm 0.189 \pm 0.981$ & $-0.750 \pm 0.153 \pm 0.122$ & $-0.735 \pm 0.240 \pm 0.413$\\
$K_2^*(1430)^0$ & $-0.084 \pm 0.052 \pm 0.161$ & $-0.119 \pm 0.046 \pm 0.161$ & $-0.176 \pm 0.066 \pm 0.102$\\
$f_2(1270)$ & $-0.354 \pm 0.047 \pm 0.076$ & $-0.402 \pm 0.051 \pm 0.041$ & $-0.279 \pm 0.067 \pm 0.115$\\
$f_2^\prime(1525)$ & $-0.827 \pm 0.175 \pm 0.259$ & $-0.757 \pm 0.167 \pm 0.281$ & $-0.764 \pm 0.263 \pm 0.438$\\
$K_3^*(1780)^0$ & $-0.936 \pm 0.080 \pm 0.740$ & $-0.934 \pm 0.069 \pm 0.661$ & $-0.971 \pm 0.036 \pm 0.165$\\
$\chi_{c0}(1P)$ & $+0.043 \pm 0.028 \pm 0.014$ & $+0.027 \pm 0.028 \pm 0.012$ & $+0.012 \pm 0.038 \pm 0.022$\\\hline
$K_0^*(1950)^0$ & $+0.050 \pm 0.097 \pm 0.059$ & $+0.014 \pm 0.081 \pm 0.034$ & ---\\\hline
$f_0(500)$ & $-0.547 \pm 0.091 \pm 0.168$ & --- & ---\\
$f_0(980)$ & $-0.054 \pm 0.017 \pm 0.015$ & --- & ---\\
$f_0(1370)$ & $-0.075 \pm 0.054 \pm 0.076$ & --- & ---\\
$f_0(1500)$ & $-0.046 \pm 0.069 \pm 0.083$ & --- & ---\\
$f_0(1710)$ & $-0.474 \pm 0.099 \pm 0.667$ & --- & ---\\
$\pi\pi$--$\pi\pi$ & $-0.079 \pm 0.049 \pm 0.069$ & --- & ---\\
$\pi\pi$--$K \Kbar$ & $+0.371 \pm 0.272 \pm 0.375$ & --- & ---\\\hline
\pip\pim S-wave & --- & $+0.052 \pm 0.033 \pm 0.109$ & --- \\\hline 
\Kp\pim and & $+0.023 \pm 0.006 \pm 0.018$ &$+0.024 \pm 0.006 \pm 0.014$ & $+0.033 \pm 0.010 \pm 0.028$ \\
\pip\pim S-waves \\\hline
Total & $+0.021 \pm 0.004 \pm 0.019$ & $+0.020 \pm 0.004 \pm 0.010$ & $+0.030 \pm 0.007 \pm 0.011$\\

    \hline
    \end{tabular}
\end{table}

\begin{table}[tb]
    \centering
    \caption{\label{tab:end:dphase}
        Quasi-two-body \CP-violating phase differences between \Bm and \Bp~($^\circ$) for the three S-wave approaches. For all measurements, the first uncertainty is statistical while the second is systematic. 
    }
    \renewcommand{\arraystretch}{1.1}
    \begin{tabular}
    {@{\hspace{0.0cm}}l@{\hspace{0.125cm}}
     @{\hspace{0.125cm}}c@{\hspace{0.125cm}}
     @{\hspace{0.125cm}}c@{\hspace{0.125cm}}
     @{\hspace{0.125cm}}c@{\hspace{0.0cm}}} \hline
    Component & Isobar & K-matrix & QMI \\ \hline

$K^*(892)^0$ & 0 (fixed) & 0 (fixed) & 0 (fixed) \\
$K^*(1410)^0$ & $\phantom{00}{-7.0} \pm \phantom{0}8.4 \pm 10.8$ & $\phantom{00}{-5.4} \pm \phantom{0}9.1 \pm \phantom{0}8.7$ & $\phantom{00}{-6.3} \pm 10.2 \pm 10.0$\\
$K^*(1680)^0$ & $\phantom{0}{-61.0} \pm 21.7 \pm 70.2$ & $\phantom{0}{-55.9} \pm 19.4 \pm 54.7$ & $\phantom{0}{-43.8} \pm 30.1 \pm 61.9$\\
$\rho(770)^0$ & $\phantom{0}{-40.3} \pm \phantom{0}5.8 \pm \phantom{0}9.6$ & $\phantom{0}{-40.3} \pm \phantom{0}6.5 \pm \phantom{0}6.0$ & $\phantom{0}{-33.6} \pm \phantom{0}8.3 \pm \phantom{0}4.5$\\
$\omega(782)$ & $\phantom{0}{-21.9} \pm 12.7 \pm \phantom{0}8.4$ & $\phantom{0}{-26.6} \pm 14.1 \pm \phantom{0}5.6$ & $\phantom{0}{-17.9} \pm 13.6 \pm \phantom{0}2.7$\\
$\rho(1450)^0$ & $\phantom{0}{-37.1} \pm 10.1 \pm 16.2$ & $\phantom{0}{-48.2} \pm 11.1 \pm \phantom{0}6.4$ & $\phantom{0}{-19.0} \pm \phantom{0}8.7 \pm \phantom{0}9.3$\\
$\rho(1700)^0$ & $\phantom{00}{+7.0} \pm 38.4 \pm 46.0$ & $\phantom{0}{+13.0} \pm 38.0 \pm 12.6$ & $\phantom{0}{+21.7} \pm 27.1 \pm 27.6$\\
$K_2^*(1430)^0$ & $\phantom{0}{-10.2} \pm \phantom{0}3.4 \pm \phantom{0}3.9$ & $\phantom{0}{-11.7} \pm \phantom{0}3.2 \pm \phantom{0}4.7$ & $\phantom{00}{-1.5} \pm \phantom{0}7.4 \pm \phantom{0}4.6$\\
$f_2(1270)$ & $\phantom{0}{+31.8} \pm \phantom{0}5.3 \pm \phantom{0}6.3$ & $\phantom{0}{+29.2} \pm \phantom{0}6.0 \pm \phantom{0}5.5$ & $\phantom{0}{+46.5} \pm \phantom{0}7.1 \pm \phantom{0}3.4$\\
$f_2^\prime(1525)$ & $-131.9 \pm 46.6 \pm 75.3$ & $-122.3 \pm 46.5 \pm 43.2$ & $-177.7 \pm 40.7 \pm 46.7$\\
$K_3^*(1780)^0$ & $\phantom{0}{-20.8} \pm 51.2 \pm 20.5$ & $\phantom{0}{-10.7} \pm 47.2 \pm 19.1$ & $\phantom{00}{+6.5} \pm 34.9 \pm 24.7$\\
$\rho_3(1690)^0$ & $\phantom{0}{-45.1} \pm 15.7 \pm 28.5$ & $\phantom{0}{-27.1} \pm 12.9 \pm 17.4$ & $\phantom{0}{-19.2} \pm 17.5 \pm 18.2$\\
$\chi_{c0}(1P)$ & $\phantom{00}{-0.4} \pm \phantom{0}7.1 \pm \phantom{0}2.6$ & $\phantom{00}{-0.2} \pm \phantom{0}6.0 \pm \phantom{0}1.6$ & $\phantom{00}{+7.3} \pm \phantom{0}7.0 \pm \phantom{0}1.3$\\\hline

$K^*_0(1430)^0$ & $\phantom{00}{-7.3} \pm \phantom{0}2.6 \pm \phantom{0}4.2$ & $\phantom{00}{-8.3} \pm \phantom{0}2.2 \pm \phantom{0}4.3$ & ---\\

$K_0^*(1950)^0$ & $\phantom{00}{-2.9} \pm \phantom{0}9.5 \pm 11.2$ & $\phantom{00}{-4.5} \pm \phantom{0}8.4 \pm \phantom{0}8.8$ & ---\\\hline
$f_0(500)$ & $\phantom{0}{-38.3} \pm \phantom{0}8.8 \pm \phantom{0}9.4$ & --- & ---\\
$f_0(980)$ & $\phantom{00}{+3.1} \pm \phantom{0}4.8 \pm \phantom{0}6.6$ & --- & ---\\
$f_0(1370)$ & $\phantom{00}{-6.1} \pm \phantom{0}4.2 \pm \phantom{0}3.4$ & --- & ---\\
$f_0(1500)$ & $\phantom{00}{-8.5} \pm \phantom{0}7.1 \pm \phantom{0}8.3$ & --- & ---\\
$f_0(1710)$ & $\phantom{0}{+35.4} \pm 13.0 \pm 41.5$ & --- & ---\\
$\pi\pi$--$\pi\pi$ & $\phantom{0}{+17.4} \pm \phantom{0}7.9 \pm 11.8$ & --- & ---\\
$\pi\pi$--$K \Kbar$ & $\phantom{0}{-71.5} \pm 30.0 \pm 25.5$ & --- & ---\\

    \hline
    \end{tabular}
\end{table}

%% file: Authorship_LHCb-PAPER-2025-069.tex
\centerline
{\large\bf LHCb collaboration}
\begin
{flushleft}
\small
R.~Aaij$^{38}$\lhcborcid{0000-0003-0533-1952},
M.~Abdelfatah$^{69}$,
A.S.W.~Abdelmotteleb$^{57}$\lhcborcid{0000-0001-7905-0542},
C.~Abellan~Beteta$^{51}$\lhcborcid{0009-0009-0869-6798},
F.~Abudin\'en$^{59}$\lhcborcid{0000-0002-6737-3528},
T.~Ackernley$^{61}$\lhcborcid{0000-0002-5951-3498},
A.A.~Adefisoye$^{69}$\lhcborcid{0000-0003-2448-1550},
B.~Adeva$^{47}$\lhcborcid{0000-0001-9756-3712},
M.~Adinolfi$^{55}$\lhcborcid{0000-0002-1326-1264},
P.~Adlarson$^{87}$\lhcborcid{0000-0001-6280-3851}
% ...

{\footnotesize \it

% ...
$^{38}$Nikhef National Institute for Subatomic Physics, Amsterdam, Netherlands\\
% ...
$^{47}$Instituto Galego de F{\'\i}sica de Altas Enerx{\'\i}as (IGFAE), Universidade de Santiago de Compostela, Santiago de Compostela, Spain\\
% ...
$^{51}$Physik-Institut, Universit{\"a}t Z{\"u}rich, Z{\"u}rich, Switzerland\\
% ...
$^{55}$H.H. Wills Physics Laboratory, University of Bristol, Bristol, United Kingdom\\
% ...
$^{57}$Department of Physics, University of Warwick, Coventry, United Kingdom\\
% ...
$^{59}$School of Physics and Astronomy, University of Edinburgh, Edinburgh, United Kingdom\\
$^{60}$School of Physics and Astronomy, University of Glasgow, Glasgow, United Kingdom\\
$^{61}$Oliver Lodge Laboratory, University of Liverpool, Liverpool, United Kingdom\\
% ...
$^{69}$Syracuse University, Syracuse, NY, United States\\
% ...
$^{87}$Department of Physics and Astronomy, Uppsala University, Uppsala, Sweden, associated to $^{60}$\\
% ...
}
\end{flushleft}